\documentclass[aoas,MSNbibl,nameyear,seceqn,dvips]{arximspdf}
\usepackage{dcolumn}
\usepackage{graphicx}

\doi{10.1214/12-AOAS561} 
\volume{6}
\issue{4}
\pubyear{2012}
\firstpage{1664}
\lastpage{1688}

\makeatletter

\newcolumntype{d}[1]{D{.}{.}{#1}}

\newtheorem{theorem}{Theorem}[section]

\newcommand{\cal}{\mathcal}

\newcommand{\dnorm}{\mathcal{N}}
\newcommand{\err}{\varepsilon}
\newcommand{\real}{\mathbb{R}}
\newcommand{\tran}{\mathsf{T}}

\newcommand{\diag}{\operatorname{diag}}

\newcommand{\eqd}{\stackrel{d}{=}}

\newcommand{\e}{\mathbb{E}}

\makeatother

\begin{document}
\begin{frontmatter}

\title{Multiple hypothesis testing adjusted for latent
variables, with an application to the AGEMAP
gene expression data}
\runtitle{LEAPP}

\begin{aug}
\author[A]{\fnms{Yunting} \snm{Sun}\thanksref{t1,t2}\ead[label=e1]{yunting@stanford.edu}},
\author[A]{\fnms{Nancy R.} \snm{Zhang}\thanksref{t2}\ead[label=e2]{nzhang@stanford.edu}}
\and
\author[A]{\fnms{Art B.} \snm{Owen}\corref{}\thanksref{t1}\ead[label=e3]{owen@stanford.edu}}
\runauthor{Y. Sun, N. R. Zhang and A. B. Owen}
\affiliation{Stanford University}
\address[A]{Department of Statistics\\
Stanford University
Sequoia Hall\\
Stanford, California 94305\\
USA\\
\printead{e1}\\
\phantom{E-mail: }\printead*{e2}\\
\phantom{E-mail: }\printead*{e3}} 
\end{aug}

\thankstext{t1}{Supported by NSF Grant DMS-09-06056.}
\thankstext{t2}{Supported by NSF Grant DMS-09-06394.}

\received{\smonth{5} \syear{2011}}
\revised{\smonth{4} \syear{2012}}

%
\begin{abstract}
In high throughput settings we inspect a great many candidate variables
(e.g., genes) searching for associations with a primary variable
(e.g., a phenotype). High throughput hypothesis testing can be made
difficult by the presence of systemic effects and other latent
variables. It is well known that those variables alter the level of
tests and induce correlations between tests.
They also change the relative ordering of significance levels among
hypotheses. Poor rankings lead to wasteful and ineffective follow-up
studies. The problem becomes acute for latent variables that are
correlated with the primary variable. We propose a two-stage analysis
to counter the effects of latent variables on the ranking of
hypotheses. Our method, called LEAPP, statistically isolates the latent
variables from the primary one. In simulations, it gives better
ordering of hypotheses than competing methods such as SVA and
\mbox{EIGENSTRAT}. For an illustration, we turn to data from the AGEMAP study
relating gene expression to age for $16$ tissues in the mouse. LEAPP
generates rankings with greater consistency across tissues than the
rankings attained by the other methods.
\end{abstract}

%

\end{frontmatter}

\section{Introduction}

There has been considerable progress in multiple
testing methods for high throughput
applications.
A common example, coming from biology, is testing which of
$N$ genes' expression levels correlate significantly with
a scalar variable, which we'll call
the primary variable.
The primary variable may be an experimentally applied treatment or it
may be a covariate
such as a phenotype.
We will use the gene expression
example for concreteness, although it is just one of
many instances of this problem.

High throughput experiments may involve
thousands or even millions of hypotheses.
Because $N$ is so large, serious problems
of multiplicity arise. For independent
tests, methods based on the false discovery
rate [\citet{dudovand2008}]
have been very successful.
Attention has turned more recently to
dependent tests [\citet{efro2010}].

One prominent cause of dependency among test statistics is the presence
of latent variables. For example, in microarray-based experiments, it
is well known that samples processed in the same batch are correlated.
Batch, technician and other sources of variation in sample preparation
can be modeled by latent variables. Another example comes from genetic
association studies, where differences in ancestral history among
subjects can lead to false or inaccurate associations. \citet
{pric2006} used principal components to extract and correct for
ancestral history, in effect modeling the genetic background of the
subjects as latent variables. A third example comes from copy number
data, where local trends along the genome cause false positive copy
number calls
[\citet{olshvenkluciwigl2004}].
\citet{disketal2008} conducted experiments showing that these local trends
correlate with the percentage of bases that are guanines or cytokines
along the genome,
and are caused by differences in the quantity and handling of DNA.
These laboratory effects are hard to measure, but can be quantified
using a latent variable model.
In this paper, we consider latent variables that might even be
correlated with the primary variable.

When the primary variable is an experimentally applied treatment,
then problematic latent variables are those that
are partially confounded with the treatment.
Randomization reduces the effects of such confounding,
but randomization is not always perfectly applied
and batch or other effects may be imbalanced with
respect to the treatment
[\citet{leekschabravsimclongjohngemabaggiriz2010}].

These latent variables have some severe consequences.
They alter the level of the hypothesis tests
and they induce correlations among multiple tests.
Another consequence, that we find especially concerning,
is that the latent variables may affect the rank ordering
among the $N$ $p$-values. When high throughput methods
are used to identify candidates for further follow-up
it is important that the highly ranked items contain
as many nonnull cases as possible.

Our approach to this problem uses a rotated
model in which we separate the latent
variables from the primary variable.
We do this
by creating two data sets, one in which
both primary and latent variables are present
and one in which the primary variables are absent.
We use the latter data set to estimate the
latent variables and then substitute their
estimates into the former. Since each gene has its own effect size in
relation to the primary variable, the former model is supersaturated.
We conduct inference under the setting where the parameter vector
relating the genes to the primary variable is sparse, as is commonly
assumed in multiple testing situations.
Each nonnull hypotheses behaves as an additive
outlier, and we then apply an outlier detection
method from \citet{sheowen2011} to find them.
We call the method LEAPP,
for \textit{l}atent \textit{e}ffect \textit{a}djustment after
\textit{p}rimary \textit{p}rojection.\vadjust{\goodbreak}

Section~\ref{secnotation} presents our notation
and introduces LEAPP along with several other related models, including
SVA [\citet{leekstor2008}] and EIGENSTRAT [\citet{pric2006}], to
which we make comparisons.
Section~\ref{secsynthetic} shows via simulation
that LEAPP generates better
rankings of the nonnull hypotheses than one would get
by either ignoring the latent
variables, by SVA, or by
EIGENSTRAT.
EIGENSTRAT estimates the latent variables (by principal
components) without first adjusting for the primary
variable. LEAPP outperforms it when
the latent variable is weaker than the primary.
EIGENSTRAT does well in simulations with
weak primary variables, which matches
the setting that motivated it. Still it is
interesting to learn that it does not extend
well to problems with strong primary variables.
SVA estimates the primary variable's coefficients
without first adjusting for correlation
between the primary and latent variables.
LEAPP outperforms it when the latent and
primary variables are correlated.

Section~\ref{secagemap} compares the methods on
the AGEMAP data of \citet{agemap}.
The primary variable there is age. While we
do not know the truly nonnull genes for this
problem, we have a proxy. The data set has
$16$ subsets, each from a different tissue type.
We find
that LEAPP gives gene lists
with much greater overlap among tissues than
the gene lists achieved by the other methods.
Our conclusions are in Section~\ref{secconclusions}.
We include some brief remarks on calibration
of the $p$-values themselves as opposed to
the rank ordering which is the primary focus
of this paper.
Some theory is given in the \hyperref[app]{Appendix}
for a simplified version of LEAPP.
The specific rotation matrix
used does not affect our answer.
For the case of one latent variable
and no covariates, the simplified LEAPP consistently
estimates the latent structure.
We also get a bound for the
sum of squared coefficient errors
when the effects are sparse.

\section{Notation and models}\label{secnotation}

In this section we describe the data model
and introduce the parameters and latent variables that arise.
Then we describe our LEAPP proposal which
is based on a series of reductions from a
heteroscedastic multivariate regression including
latent factors to
a single linear regression problem with additive
outliers and known error variance.
We also describe EIGENSTRAT and SVA,
to which we make comparisons, and then survey
several other published methods for this problem.

\subsection{Data, parameters, latent variables and tests}\label{secvars}

The data we observe are
a response matrix $Y\in\real^{N\times n}$
and a variable of interest $g\in\real^n$,
which we call the primary variable.
In an expression problem $Y_{ij}$ is the
expression level of gene $i$ for subject $j$.
Very often the primary variable $g$ is a group
variable taking just two values, such
as $\pm1$ for a binary phenotype,
then linearly transformed to have
mean~$0$ and norm $1$.
The quantity $g_j$ can also be a more general
scalar, such as the age of subject $j$.

We are interested to know which genes,
if any, are linearly associated with the variable $g$.
We capture this linear association\vadjust{\goodbreak}
through the $N\times n$ matrix $\gamma g^\tran$, where
$\gamma$ is a vector of $N$ coefficients.
When most genes are not related
to $g$, then $\gamma$ is sparse.

Often there are covariates $X$ other than $g$
that we should adjust for. The covariate
term is $\beta X^\tran$
where $\beta$ contains coefficients.
The latent variables that cause tests
to be mutually correlated are assumed to take
an outer product form $UV^\tran$. Neither
$U$ nor $V$ is observed.
Finally, there is observational noise
with a variance that is allowed to be
different for each gene, but assumed to be constant
over subjects.

The full data model is
%
\begin{equation}
\label{eqfull} Y = \gamma g^\tran+ \beta X^\tran+
UV^\tran+\Sigma E
\end{equation}
for variables
\begin{eqnarray*}
Y &\in& \real^{N\times n} \qquad\mbox{response values},
\\
g &\in& \real^{n\times1} \qquad\mbox{primary predictor, that is, treatment, with
$g^\tran g=1$},
\\
\gamma&\in& \real^{N\times1} \qquad\mbox{primary parameter, possibly sparse},
\\
X &\in& \real^{n\times s} \qquad\mbox{$s$ covariates (e.g., sex) per
subject},
\\
\beta&\in& \real^{N\times s} \qquad\mbox{$s$ coefficients, including per gene
intercepts},
\\
U &\in& \real^{N\times k} \qquad\mbox{latent, nonrandom rows (e.g., genes)},
\\
V &\in& \real^{n\times k} \qquad\mbox{latent, independent rows (e.g.,
subjects)},
\\
E &\sim& \dnorm(0,I_N
\otimes I_n) \qquad\mbox{noise}
\end{eqnarray*}
and
\[
\Sigma = \diag(\sigma_1,\ldots,\sigma_N)\qquad
\mbox{standard deviations}
\]
with dimensions
\begin{eqnarray*}
&&n \qquad\mbox{number of arrays/subjects},
\\
&&N \gg n \qquad\mbox{number of genes},
\\
&&s \ll n \qquad\mbox{number of covariates}
\end{eqnarray*}
and
\[
k \ge 1 \qquad\mbox{latent dimension.}
\]

After adjusting for $X$,
the genes are correlated through the action of
the latent portion $UV^\tran$ of the model.
They may have unequal variances, through both
$\Sigma$ and $U$. We adopt the normalization
$\e(V^\tran V)=I_k$.
It is possible to generalize the model to have
a primary variable $g$ of dimension
larger than one,
but we focus on the case of a single primary variable.

We pay special attention to the case of $k=1$
latent variable.
The algorithm is the same for all values of $k$.
But, when $k=1$, the dependence
between the variable $g$ of interest
and the latent variable $V$ can be summarized
by a single correlation coefficient $\rho=g^\tran V/\sqrt{V^\tran V}$
which aids interpretation.\vadjust{\goodbreak}

Writing (\ref{eqfull}) in terms of indices yields
%
\begin{equation}
\label{eqindices} Y_{ij} = \gamma_i g_j +
\beta_i^\tran X_j + U_i^\tran
V_j +\sigma_i \err_{ij},\qquad 1\le i\le N, 1\le j
\le n.
\end{equation}
Here $\beta_i$ and $U_i$ are
the $i$th rows of $\beta$ and $U$, respectively,
as column vectors.
Similarly, $X_j$ and $V_j$ are the $j$th rows
of $X$ and $V$, $\sigma_i$ is the $i$th diagonal
element of $\Sigma$ and $\err_{ij}$ is the $ij$
element of $E$.

Our LEAPP proposal is based on a series of
reductions described next. In outline,
we first split the data into two parts, one of which
is completely free of the primary variable. We then
estimate some properties of the latent variable
model from that primary-free data. Finally, we use
those estimated quantities in the part of the data which
does contain the primary variable to identify
genes related to the primary variable.

\subsection{Data rotation}
We begin by choosing an orthogonal matrix $O\in\real^{n\times n}$
such that $g^\tran O^\tran
= (\eta,0,0,\ldots,0)\in\real^{1\times n}$
where $\eta= \Vert g\Vert>0$.
Without loss of generality, we assume
that the primary
predictor has been scaled so that $\eta=1$.
A~convenient choice for $O$ is the
Householder matrix $O=I_n-2\kappa\kappa^\tran$,
where $\kappa= (g-e_1)/\Vert g-e_1\Vert_2$
and $e_1=(1,0,\ldots,0)^\tran$.

Using $O$, we construct the \textit{rotated} model
%
\begin{eqnarray}
\label{eqrotated} 
Y^{(r)}&\equiv& YO^\tran = \gamma
g^\tran O^\tran+ \beta X^\tran O^\tran+
UV^\tran O^\tran+\Sigma E O^\tran
\\
&\equiv& \gamma g^{(r)\tran} + \beta X^{(r)\tran} +
UV^{(r)\tran}+\Sigma E^{(r)},
\end{eqnarray}
where $g^{(r)}$, $X^{(r)}$, $V^{(r)}$ and
$E^{(r)}$ are rotated versions of
$g$, $X$, $V$ and $E$, respectively.
For each major transformation of the data, a
new mnemonic superscript will be introduced.
Some superscripts use the same letter also
used as a data dimension, but the usages are
distinct enough that one will not be mistaken
for the other.

Notice that $E^{(r)}=EO^\tran\eqd E$,
because $E \sim\dnorm(0, I_N\otimes I_n)$.
By construction, $g^{(r)}= (1,0,\ldots,0)$.
Therefore, the model for $Y^{(r)}_{ij}$
is different depending on whether $j=1$ or $j\ne1$:
%
\begin{equation}
\label{eqdriver}
Y^{(r)}_{i1}  = \beta_i^\tran
X^{(r)}_1 + U_i^\tran
V^{(r)}_1+ \gamma_i + \sigma_i
\err^{(r)}_{i1}
\end{equation}
and
\begin{equation}
\label{eqrest}
Y^{(r)}_{ij} = \beta_i^\tran
X^{(r)}_j + U_i^\tran
V^{(r)}_j+ \sigma_i\err^{(r)}_{ij},\qquad
j=2,\ldots,n,
\end{equation}
where $\err^{(r)}_{ij}$ is the $(i,j)$th element of $E^{(r)}$.

The rotated model concentrates the primary coefficients $\gamma_i$
in the first column of~$Y^{(r)}$.
Our approach is to base tests and
estimates of $\gamma_i$ on equation (\ref{eqdriver}).
We need to substitute estimates for
unknown quantities $\sigma_i$, $\beta_i$
and $U_i$ in (\ref{eqdriver}).
The estimates come from the
model in equation (\ref{eqrest}).

This rotated approach has some practical
advantages:
First,
we do not need to iterate between
applying equations (\ref{eqdriver}) and (\ref{eqrest}).
Instead we use (\ref{eqrest}) once
to estimate unknowns $U$, $\sigma$ and $\beta$
and then use (\ref{eqdriver}) once
to judge $\gamma_i$.
Second,\vspace*{2pt} the last $n-1$ columns
of $Y^{(r)}$, and hence estimates $\widehat\sigma$,
$\widehat\beta$, and $\widehat U$,
are statistically independent of the first column.
Third, problems (\ref{eqdriver}) and (\ref{eqrest})
closely match settings for which there are usable methods
as described next.

Using estimates $\widehat{\sigma}_i$, $\widehat{U}_i$ and $\widehat{\beta}_i$
from (\ref{eqrest}) described below,
we may write (\ref{eqdriver}) as
%
\begin{equation}
\label{eqreviseddriver} Y^{(r)}_{i1}-\widehat{
\beta}_i^\tran X^{(r)}_1 = \widehat
U{}^\tran_i V^{(r)}_1 +
\gamma_i +\widehat\sigma_i\err^{(r)}_{i1}.
\end{equation}
The right-hand side of
equation (\ref{eqreviseddriver}) is a regression
with measurement errors in the predictors $\widehat U_i$,
mean-shift outliers $\gamma_i$ and unequal error variances.
We will use the $\Theta$--IPOD algorithm
of \citet{sheowen2011}, adjusted to
handle unequal $\sigma_i$, to get our estimate of $\gamma_i$.

Before describing $\Theta$--IPOD we show
how to get the estimates $\widehat\beta_i$,
$\widehat U_i$ and $\widehat\sigma_i$ from the
criss-cross regression algorithm of \citet{gabrzami1979}.
Criss-cross regression will also produce an estimate of $V^{(r)}_j$
for $j\ge2$, but those vectors do not play a role in (\ref{eqreviseddriver}).

\subsection{\texorpdfstring{Estimating $U$, $\beta$ and $\Sigma$}{Estimating U, beta and Sigma}}
We get our estimates of $U_i$, $\beta_i$ and $\sigma_i$
from the last $n-1$ columns of the data set.
Let $Y^{(\ell)}$, $X^{(\ell)}$, $V^{(\ell)}$ and $E^{(\ell)}$ be
the \textit{last}
$n-1$ columns of $Y^{(r)}$, $X^{(r)}$, $V^{(r)}$ and $E^{(r)}$, respectively.
Then the model for the last $n-1$ columns of the data is
%
\begin{equation}
\label{eqlastpart} Y^{(\ell)}= \beta X^{(\ell)\tran} + U
V^{(\ell)\tran} + \Sigma E^{(\ell)}.
\end{equation}
Notice that the quantities $\beta$, $U$ and $\Sigma$
in (\ref{eqlastpart}) are the same as
those in the original model (\ref{eqfull}) because
the steps taken so far operate on columns of~$Y$.
We can write $Y^{(\ell)}=Y^{(r)}D_n$ where
$D_n={0\choose I_{n-1}}\in\real^{n\times(n-1)}$
and similarly for $X^{(\ell)}$ and $V^{(\ell)}$.
The matrix $D_n$ deletes the first column
out of $n$ in the matrix that it follows.

We adopt an iterative approach based on (\ref{eqlastpart})
that alternates between updating $\widehat\Sigma$
and updating the quantities $\widehat\beta$, $\widehat U$ and ${\widehat V}^{(\ell
)}$ given $\widehat\Sigma$.
The update for $\widehat\Sigma$ is
%
\begin{equation}
\widehat\Sigma= \biggl(\frac1{n-1}\diag\bigl( \widehat\err\widehat\err^\tran\bigr)
\biggr)^{1/2}\qquad \mbox{where } 
\widehat\err= Y^{(\ell)}- \widehat
\beta X^{(\ell)}-\widehat U 
{{ {\widehat V}}^{(\ell)\tran}}.
\end{equation}
That is, $\widehat\sigma^2_i$ is simply the mean squared
error of a regression for the $i$th gene.

Given $\widehat\Sigma$, we \textit{standardize} the last $n-1$
columns, yielding $Y^{(s\ell)} = \widehat\Sigma^{-1}Y^{(\ell)}$.
In terms of the other variables,
%
\begin{equation}
\label{eqprecriss} Y^{(s\ell)} = \beta^{(s)} X^{(\ell)\tran} +
U^{(s)} V^{(\ell
)\tran} + E^{(s\ell)},
\end{equation}
where
$\beta^{(s)}=\widehat\Sigma^{-1}\beta$,
$U^{(s)} = \widehat\Sigma^{-1}U$ and $E^{(s\ell)}=\widehat\Sigma^{-1}E^{(\ell)}$
are standardized versions of $\beta$, $U$ and $E^{(\ell)}$,
respectively.

Because $\Sigma^{-1}E^{(\ell)}$ has IID Gaussian entries,
equation (\ref{eqprecriss}) closely matches
the criss-cross regression model
of \citet{gabrzami1979}.
Criss-cross regression for a matrix of data
sums three outer products:
row based features (with column coefficients),
column based features (with row coefficients),
and a low rank factor model with latent
rows and columns.

We fit a criss-cross regression by first
estimating $\beta^{(s)}$ by least
squares regression:
\[
\widehat\beta^{(s)} = {Y^{(s\ell)}}X^{(\ell)}
\bigl(X^{(\ell)\tran} X^{(\ell)}\bigr)^{-1}.
\]
Then\vspace*{1pt} we estimate
$U^{(s)}$ and $V^{(\ell)}$ by a truncated
singular value decomposition (SVD)
of rank $k$ applied to the residuals
$\widehat\err^{(s\ell)}=Y^{(s\ell)}-\widehat\beta^{(s)}
X^{(\ell)\tran}$.
We absorb the singular values into $\widehat U^{(s)}$
but retain the identity
${\widehat V}^{(\ell)\tran} {\widehat V}^{(\ell)}=I_k$.

Our use of criss-cross regression has
a latent factor model of the form
$UV^\tran$ and terms
of the form $\beta X^\tran$ representing
column features with row coefficients.
The full criss-cross regression model
also allows for terms of the form $Z\delta^\tran$
that combine row features with column coefficients.

To apply the algorithm,
we need a starting point for the iteration
and a value of~$k$.
We start with $\widehat\Sigma= I_N$.
We have assumed that the rank $k$ for the
latent variables is known. When it must
be estimated from the data,
we follow \citet{leekstor2008} in using
the method of \citet{bujaeyob1992}, as described
in Section~\ref{secsva}.

Criss-cross\vspace*{1pt} regression gives us
estimates $\widehat\Sigma$, $\widehat\beta^{(s)}$ and
$\widehat U^{(s)}$.
We can estimate
$\widehat\beta$ by $\widehat\Sigma^{1/2}\widehat\beta^{(s)}$
and $\widehat U$ by $\widehat\Sigma^{1/2}\widehat U^{(s)}$.
We will use these estimates normalized by $\widehat\sigma_i$
and so it is also possible to work
with $\widehat\beta^{(s)}$ and $\widehat U{}^{(s)}$ themselves.

\subsection{Gene identification}

Now we return to the first column of the
rotated data matrix which contains the effects
of the primary variable.
If we divide $Y^{(r)}_{i1}$ by $\sigma_i$, we get
%
\begin{equation}
\label{eqprimarystd} \frac{Y^{(r)}_{i1}}{\sigma_i} = \frac{\beta
_i^\tran}{\sigma_i} X^{(r)}_1
+ \frac{U_i^\tran}{\sigma_i} V^{(r)}_1+ \frac{\gamma_i}{\sigma_i} +
\err^{(r)}_{i1},\qquad i=1,\ldots,N.
\end{equation}
For our purposes, equation (\ref{eqprimarystd}) can be cast
as a regression of standardized variables on $k$ predictors
$U_i/\sigma_i$ with coefficient vector $V_1^{(r)}\in\real^k$,
with additive outliers $\gamma_i/\sigma_i$ and
offsets $\beta_i^\tran X_1^{(r)}/\sigma_i$.
Though $\sigma_i$ and $\beta_i$ and $U_i$
are unknown, we have estimates of them
from the previous section.

We use those estimates to construct the \textit{primary}
variable regression model
%
\begin{equation}
\label{eqprimarymodel} Y_i^{(p)} = U_i^{(p)\tran}
V_1^{(p)} +\gamma_i^{(p)}+
\err^{(p)}_i
\end{equation}
with response
$Y_i^{(p)} = (Y_{i1}^{(r)}-\widehat\beta_i^\tran X_1^{(r)})/\widehat\sigma_i$,
predictors $U_i^{(p)} = \widehat U_{i1}^{(r)}/\widehat\sigma_i = \widehat U_{i1}^{(s)}$,
coefficient vector $V_1^{(p)}=V_1^{(r)}$, additive outliers
$\gamma_i^{(p)} =\gamma_i/\widehat\sigma_i$, and error
$\err_i^{(p)} =
\err_{i1}^{(r)}\sigma_i/\widehat\sigma_i$.

The $\Theta$--IPOD algorithm of \citet{sheowen2011}
is designed to estimate a regression coefficient
in the presence of additive outliers as well as
to identify which observations are outliers.
In the present context, the outliers
correspond to genes that are associated with the
primary variable.

For a complete description of $\Theta$--IPOD
see \citet{sheowen2011}, who also cite related
work in the robust regression literature. Here we give
a brief account of the main points.

The primary variable model (\ref{eqprimarymodel})
could be fit by minimizing
$\Vert Y^{(p)} - U^{(p)} V_1^{(p)}\Vert_2^2 + \lambda\Vert
\gamma^{(p)}\Vert$ over $V_1^{(p)}$ and $\gamma^{(p)}$.
Large enough penalties $\lambda>0$ would yield
a sparse estimate of $\gamma^{(p)}$ which is
desirable because the model has $N+k$
parameters and only $N$ observations.

The natural algorithm to minimize the sum of squared errors with an
$L_1$ penalty on the additive outlier coefficients
alternates between two steps. One step estimates
the additive outlier effects by soft thresholding
residuals from a least squares regression. The other
step does the least squares regression after first
subtracting the estimated outlier effects.
\citet{sheowen2011} found that while soft thresholding
is not robust, simply changing the algorithm to do
hard thresholding proved to be very robust. Their
algorithm also takes account of the leverage values
in least squares regression. The algorithm
requires a choice for $\lambda$.
They used a modified BIC statistic from \citet{chenchen2008}.

Our statistic for testing $H_{i0}\dvtx\gamma_i=0$ is
%
\begin{equation}
\label{Ti} T_i = \frac{Y_i^{(p)}- U_i^{(p)\tran}\widehat V_1}{\widehat\tau},
\end{equation}
where $\widehat V_1$ is the $\Theta$--IPOD estimate
of $V_1^{(p)}$ and $\widehat\tau$
is an estimate of the error variance from (\ref{eqprimarymodel}).
The estimate $\widehat\tau$ is
the median absolute deviation
from the median (MAD) of ${Y_i^{(p)}-
U_i^{(p)\tran}\widehat V_1}$,
with the customary scaling
to match the standard deviation for a Gaussian
distribution.

For $p$-values we
use $\Pr( |Z|\ge|T_i|)$ where $Z\sim\dnorm(0,1)$.
Candidate hypotheses are ranked from most interesting
to least interesting by taking the $p$-values
from smallest to largest.
This is equivalent to sorting $|T_i|$ from largest
to smallest. We consider the
quality of this ordering, not whether the
$p$-values are properly calibrated, apart from a brief
remark in the conclusions.

The entire LEAPP algorithm is summarized in
Figure~\ref{figleapp}.

\begin{figure}
\renewcommand\theenumi{(\arabic{enumi})}
\renewcommand\labelenumi{\theenumi}
\begin{enumerate}[(14)]
\item Standardize the primary variable, $g=g/\Vert g\Vert$.
\item Define the rotation matrix $O=I_n-2\kappa\kappa^\tran$
for $\kappa=(g-e_1)/\Vert g-e_1\Vert$.
\item Rotate $Y^{(r)}=YO^\tran$ and $X^{(r)}=XO^\tran$.
\item Select the last $n-1$ columns $Y^{(\ell)}=Y^{(r)}D_n$ and
$X^{(\ell)}=X^{(r)}D_n$.
\item Let $\widehat\beta^{(s)} = Y^{(\ell)\tran} X^{(\ell)}
(X^{(\ell)\tran} X^{(\ell)})^{-1}$.
\item \label{stepbuja} Use \citet{bujaeyob1992} to estimate
the rank $k$ for $Y^{(\ell)}-\widehat\beta^{(s)} X^{(\ell)}$.
\item Set $\widehat\Sigma=I_N$.
\item\label{stepcrisscross} Iterate to convergence:
\end{enumerate}
{\renewcommand\theenumi{(\alph{enumi})}
\renewcommand\labelenumi{\theenumi}
\begin{enumerate}[(14)\mbox{ }\hspace*{3pt}(a)]
\item $Y^{(s\ell)} = \widehat\Sigma^{-1} Y^{(\ell)}$.
\item $\widehat\beta^{(s)} = Y^{(s\ell)\tran} X^{(\ell)}
(X^{(\ell)\tran} X^{(\ell)})^{-1}$.
\item $\widehat E^{(s\ell)}_k$ gets rank $k$ truncated SVD of $\widehat
E^{(s\ell)}
= {Y^{(s\ell)}}- \widehat\beta^{(s)} X^{(\ell)\tran}$.
\item \label{stepstepsighat}
$\widehat\Sigma=
(\diag(
(\widehat E^{(s\ell)}-\widehat E^{(s\ell)}_k)
(\widehat E^{(s\ell)}-\widehat E^{(s\ell)}_k)^\tran
)/(n-1) )^{1/2}$.
\end{enumerate}}
\begin{enumerate}[(14)]
\setcounter{enumi}{8}
\item Let $\widehat U^{(s)}$ be the $k$ right singular vectors of $\widehat
E^{(s\ell)}$.
\item Set $\widehat\beta= \widehat\Sigma\widehat\beta^{(s)}$, $\widehat U =
\widehat\Sigma\widehat U^{(s)}$.
\item Set $Y_i^{(p)}=(Y_{i1}^{(r)}-\widehat\beta_i^\tran
X_1^{(r)})/\widehat\sigma_i$,
$U_i^{(p)} = \widehat U_i^{(s)}$.
\item Fit $\Theta$--IPOD with response $Y_i^{(p)}$ predictors $U_i^{(p)}$
getting $\widehat\gamma_i^{(p)}$, $\widehat V_1^{(p)}$ and $\widehat\tau$.
\item Let $T_i = (Y_i^{(p)} -
U_i^{(p)\tran}{\widehat V}_1^{(p)})/\widehat\tau$, $i=1,\ldots,N$.
\item Rank genes from most significant (largest $|T_i|$) to least.
\end{enumerate}
\caption{The LEAPP algorithm, using notation from the text.
Step \protect\ref{stepbuja} can be omitted if the desired value of
$k$ is already known. Step (8)\protect\ref{stepstepsighat} is
written concisely but can be computed more efficiently.
We use $|T_i|$ to rank genes.
Convergence at \protect\ref{stepcrisscross} is
declared when
$\Vert\widehat\Sigma_{\mathrm{new}}-\widehat\Sigma_{\mathrm{old}}\Vert_1
/\Vert\widehat\Sigma_{\mathrm{old}}\Vert_1< 10^{-4}$
with $\Vert\cdot\Vert_1$ here being the sum of absolute
diagonal elements.
There is an R package for LEAPP
at \protect\url{http://cran.r-project.org/web/packages/leapp/}.%
}\label{figleapp}
\end{figure}

We have emphasized the setting in which $\gamma$
is a sparse vector. When $\gamma$ is not a sparse
vector, then its large components may not
be flagged as outliers because the MAD estimate of $\tau$ would be
inflated due to contamination by $\gamma$. In this case, however,
we can fall back on a simpler approach to estimating~$\tau$.
The error $\err^{(p)}_i$ has variance
$\e(\sigma^2_i/\widehat\sigma_i^2 )$.
This variance differs from unity only because
of estimation errors in $\widehat\sigma_i$.
We can then use $\tau^2=1$.
We can account for fitting $s$ regression parameters to
the $n-1$ samples in each row of $Y^{(\ell)}$ by taking
$\tau^2 = \e((n-1-s)/\chi^2_{n-1-s})
=(n-s-1)/(n-s-3)$.
A further approximate adjustment for estimating
$k$ latent vectors is to take
$\tau^2 =(n-s-k-1)/(n-s-k-3)$. This estimate of $\tau$ can be used in
(\ref{Ti}) for ranking of hypotheses if $\gamma$ is not suspected to
be sparse.

\subsection{SVA}\label{secsva}

We compare our method to
the surrogate variable analysis
(SVA) method of \citet{leekstor2008}.
Their iteratively reweighted surrogate variable
analysis algorithm
adjusts for latent variables before doing a regression.
But it does not isolate them.

A full and precise description of SVA appears in
the supplementary information
and online software for \citet{leekstor2008}.
Here we present a brief outline.
Their model takes the form
\[
Y = \gamma g^\tran+ UV^\tran+ \Sigma E,
\]
where $UV^\tran$ is their ``dependence kernel''
and $E$ is not necessarily normally distributed
but has independent rows.

The SVA algorithm uses iteratively reweighted SVDs
to estimate $U$, $V$ and~$\gamma$. The weights
are empirical Bayes estimates of $\Pr(\gamma_i=0, U_i\ne0\mid
Y,g,V)$ from \citet{storakeykrug2005}.
Their method seeks to remove the primary
term $\gamma g^\tran$
by downweighting rows with $\gamma_i\ne0$.
Our method creates columns that are free of the primary
variable by rotation.

The SVA iteration is as follows.
First, they fit a linear model without any latent variables, getting
estimates $\widehat{\gamma}$ and the residual
$R = Y-\widehat{\gamma}g^\tran$. Second, they apply the simulation method
of \citet{bujaeyob1992} to $R$ to estimate the number $k$ of factors,
and then take the top $k$ right eigenvectors of $R$ as the initial
estimator $\widehat{V}$. Third, they form the empirical Bayes
estimates $w_i=\Pr(\gamma_i=0, U_i\ne0\mid Y,g,\widehat{V})$
from \citet{storakeykrug2005}. Fourth, based on those weights,
they perform a weighted singular value decomposition of
the original data matrix $Y$, where row
$i$ is weighted by $w_i$.
The weighted SVD gives them an updated estimator $\widehat{V}$.
They\vspace*{1pt} repeat steps (3) and (4),
revising the weights $w_i$ and then the
matrix $\widehat{V}$, until $\widehat{V}$ converges.
They perform significance analysis on $\gamma$ through
the multivariate linear regression model
\[
Y = \gamma g^\tran+ U\widehat{V}{}^\tran+\Sigma E,
\]
where $\widehat{V}$ is treated as known
covariates to adjust for the primary effect $g$.

To estimate the
number $k$ of factors in the SVD,
they use a simulation method of \citet{bujaeyob1992}.
That algorithm uses Monte Carlo sampling to adjust
for the well-known problem that the largest singular
value in a sample covariance matrix is positively biased.
That method has two parameters: the number of simulations
employed and a significance threshold.
The default significance
threshold was $0.1$ and the default uses
$20$ permutations.

\subsection{EIGENSTRAT}

EIGENSTRAT [\citet{pric2006}] was developed to control for differences
in ancestry in genetic association studies, where the matrix $Y$
represent the alleles carried by the subjects at the genetic markers
(e.g., $Y_{ij} \in\{0,1,2\}$ counts the number of one of the
alleles). The primary variable can be case versus control, disease
status or other clinical traits.

In our notation,
they begin with a principal components
analysis approximating $Y$ by $\widehat U\widehat V^\tran$
for $\widehat U\in\real^{N\times k}$
and $\widehat V\in\real^{n\times k}$.
Then for $i=1,\ldots,N$ they
test whether $Y_{i,1: n}$ is significantly
related to $g$ in a regression including
the $k$ columns of $\widehat V$ or, equivalently,
whether the partial correlation of $Y_{i,1: n}$
on $g$, adjusted for $\widehat V$, is significant.\vadjust{\goodbreak}
Although the data are discrete and the method
resembles one for Gaussian data, the
results still clearly obtain latent variables
showing a natural connection to the geographical
region of the subjects' ancestors.

EIGENSTRAT\vspace*{1pt} has an apparent weakness. If the
signal $\gamma g^\tran$ is large, then its presence will
corrupt the estimates of $\widehat U$ and $\widehat V$.
The estimate $\widehat V$ will be correlated with
the effect $g$ that we are trying to estimate
a coefficient for.
Indeed, we find in our simulations of
Section~\ref{secsynthetic} that EIGENSTRAT
performs poorly when the signal is large compared
to the latent variable.
While EIGENSTRATs strong latent with weak signal assumption seems to
be appropriate for genetic association studies,
a method that does not rely on such assumptions is desirable.

EIGENSTRAT also requires the choice of
a rank $k$ for the latent term.
\citet{pric2006} describe a
default choice of $k=10$.
\citet{pattpricreic2006} apply
a spiked covariance model test of
\citet{john2001} using the
Tracy--Widom distribution [\citet{tracwido1994}].

\subsection{Other methods}

We have used Eigenstrat and SVA in our comparisons
because they are widely used in applications.
A number of other methods have been proposed
for this problem.
It is not feasible to include them all
in our numerical comparisons.
Instead we describe several of them here,
relating their approaches to the notation
of Section~\ref{secvars}.

\citet{frigkloacasu2009}
model their data as $Y=\gamma g^\tran+ UV^\tran+\Sigma E$.
They assume the latent $V$ is normally distributed
(independent of $E$) and that $U$ is nonrandom.
They do not assume sparsity for $\gamma$.
They estimate $U$, $V$, $\gamma$ and $\Sigma$
by an EM algorithm.
They find that using $\widehat V$ in an FDR
procedure
is an improvement compared to a model that does not employ latent variables.

\citet{lucakungchi2010}
take $Y=\beta X^\tran+ UV^\tran+\Sigma E$
and make extensive use of sparsity priors.
They include the primary variable $g$
as one of the columns of $X$, instead of
singling it out as we do.
Under their sparsity priors, a coefficient is either
$0$ or it is $\dnorm(0,\tau^2)$. The probability of a nonzero
coefficient is $\pi$, which in turn has a Beta distribution
with a small mean. They apply sparsity priors to
the elements of both the coefficient matrix
$\beta$ and the latent variables $U$. The parameters $\pi$
and $\tau$ are different for each column of $\beta$.
They use Markov chain Monte Carlo for their inferences.

\citet{alletibs2010tr}
model the data as $Y=\gamma g^\tran+E$
where $E\sim\dnorm(0,\Sigma\otimes\Gamma)$.
That is, the noise covariance is of Kronecker
form which models dependence between rows and
between columns. Our model has a different
variance equal to the sum of two Kronecker
matrices, one from $UV^\tran$ and one from $\Sigma E$.
They estimate their $\Sigma$
and $\Gamma$ by maximum likelihood with
a penalty on the norm of the inverses
of $\Sigma$ and $\Gamma$. Their $L_1$
penalties encourage sparsity in $\widehat\Sigma^{-1}$
and $\widehat\Gamma^{-1}$.
They then whiten $Y$ using $\widehat\Gamma$ and $\widehat\Sigma$
and apply false discovery rate methods.\vadjust{\goodbreak}
They also show that correlations among different columns lead to
incorrect estimates of FDR, while correlated rows do not much affect
the estimates of FDR.

\citet{efro2007} proposed a method to fit an empirical null to the
data to directly account for correlations across arrays.
The empirical null method works with estimated $Z$
scores (one per gene) and uses the histogram of those
scores to account for the effects of latent variables.
This process adjusts significance levels for hypotheses
but does not alter their ordering.

\citet{carvetal2008} consider similar problems but
apply a very different formulation. They treat
the primary variable (our $g$) as the response
and use the data matrix (our $Y$) as predictors.

\subsection{Rank estimation}

The problem of choosing the number $k$
of latent variables is a difficult one
that arises for all the methods we used.
The Tracy--Widom strategy is derived
for the case with $\Sigma= \sigma I_N$, while
our motivating applications have heteroscedasticity.

Even for $\Sigma=\sigma I_N$ it is known that
the best rank for estimating
$UV^\tran$ is not necessarily the true rank.
There is a well-known threshold strength below which
a factor is not detectable and
\citet{perr2009} shows that there is a still
higher threshold below which estimating
that factor worsens the estimate of $UV^\tran$.
\citet{bicros} present a cross-validatory
estimate for the rank $k$ and
\citet{perr2009} shows how to tune it
to choose a rank $k$ that gives the
best reconstruction as measured by the Frobenius
norm.

In our numerical comparisons, LEAPP, SVA and EIGENSTRAT
were all given the same rank $k$
to use. Sometimes $k$ was fixed at
a default value. Other times
we used the method of \citet{bujaeyob1992}.

\section{Performance on synthetic data}\label{secsynthetic}

In this section we generate data from the model (\ref{eqfull})
and compare the results from the algorithms
to each other, to an oracle which is
given the latent variable, and to
a raw regression method which makes no
attempt to adjust for latent variables.
Some simulations by \citet{sun2011} made
under a different model are described in Section~\ref{secconclusions}.

We choose $s=0$, omitting the $\beta X^\tran$
covariate term, so the simulated data satisfy
%
\begin{equation}
\label{eqsimu} Y = \gamma g^\tran+ UV^\tran+\Sigma E.
\end{equation}
The model (\ref{eqsimu}) is a special case of both
the LEAPP model and the SVA model.

Our simulations have $n=60$ (subjects) and $N=1000$
(genes).
Our primary covariate is a binary treatment
vector $g\propto(1,\ldots,1,-1,\ldots,-1)$, with
equal numbers of $1$ and $-1$, normalized
so that $g^\tran g=1$.

The vector $\gamma$ of treatment effects
has independent components $\gamma_i$
taking the values $c>0$ and $0$ with
probability $\pi=0.1$ and $1-\pi=0.9$, respectively.
We chose $c$ in order to attain
specific signal to noise ratios as described
below.
The matrix $\Sigma$ is a diagonal\vadjust{\goodbreak}
with nonzero entries $\sigma_i$
sampled independently from an
inverse gamma distribution:
$1/\sigma^2_i\sim\operatorname{Gamma}(5)/4$.
Note that $\e(\sigma^2_i)=1$.

We use $k=1$ latent variable that has
correlation $\rho$ with $g$.
The latent vector $U=(u_1,\ldots,u_N)$
is generated as independent $U(-a,a)$
random variables. We will choose $a$
to obtain specific latent
to noise variance ratios.
The latent vector $V$ is taken to be
$\rho g +\sqrt{1-\rho^2} W$,
where $W$ is uniformly distributed on the
set of unit vectors orthogonal to $g$.
That is, we sample $V$ so as to have
a sample correlation and squared norm
that both match their population counterparts.

The model (\ref{eqsimu})
gives $Y$ three components:
the signal ${\cal S}=\gamma g^\tran$,
the latent structure ${\cal L}=UV^\tran$,
and the noise ${\cal N}=\Sigma E$.
The relative sizes of these components
affect the difficulty of the problem.
We use Frobenius and spectral norms
to describe the sizes of these matrices.

The noise matrix is constructed
so that $\e( \sigma^2_i\err_{ij}^2)
=\e(\sigma^2_i)=1$, so that
$\e(\Vert{\cal N}\Vert_F^2)=Nn$.
Because the signal and latent matrices
have rank $1$,
%
\begin{equation}
\label{eqsignorm}
\e\bigl(\Vert{\cal S}\Vert_F^2\bigr) = \e\bigl(\Vert{
\cal S}\Vert_2^2\bigr)=\e\bigl(\Vert\gamma
\Vert^2_2\bigr) = N\pi c^2
\end{equation}
and
\begin{equation}
\label{eqlatnorm}
\e\bigl(\Vert{\cal L}\Vert_F^2\bigr) = \e\bigl(\Vert{
\cal L}\Vert_2^2\bigr)=\e\bigl(\Vert U
\Vert^2_2\bigr) = Na^2/3.
\end{equation}
For our simulation, we specified the ratios
\[
\mbox{SNR} \equiv\pi c^2 \quad\mbox{and}\quad \mbox{LNR} \equiv
a^2/3
\]
and varied them over a wide range.
We also use $\mbox{SLR}=3\pi c^2/a^2$.

We also varied the level of $\rho$, the correlation
between the latent and primary variables.
For each setting of SNR, SLR, LNR and $\rho$
under consideration, we simulated the process $100$
times and prepared ROC curves, from the pooled
collection of $100{,}000$ predictions.

The methods that we applied are as follows:\vspace*{8pt}

\begin{tabular}{ll}
\textit{true} &
an oracle given $UV^\tran$ which then does regression of $Y-UV^\tran$
on $g$,\\
\textit{raw} & multivariate regression of $Y$ on $g$ ignoring latent
variables,\\
\textit{eig} & EIGENSTRAT of \citet{pric2006},\\
\textit{sva} & surrogate variable analysis from \citet{leekstor2008},
and\\
\textit{lea} & our proposed LEAPP method.
\end{tabular}\vspace*{8pt}

\begin{figure}

\includegraphics{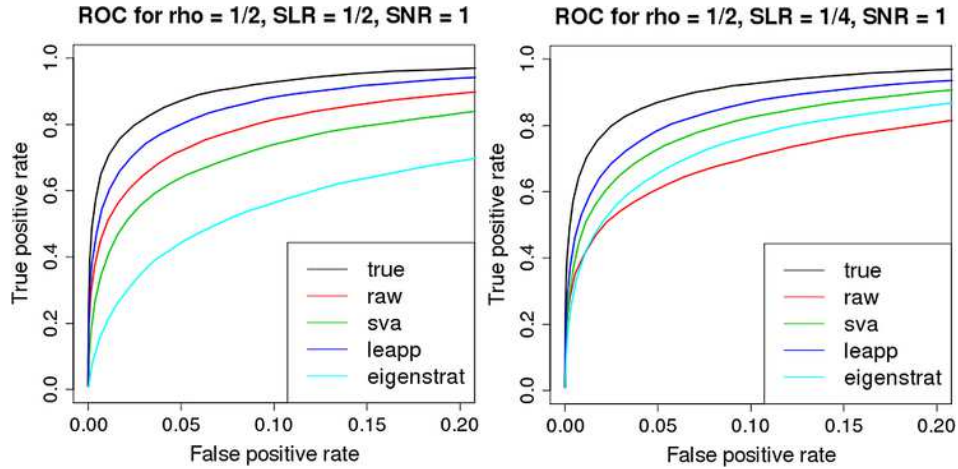}

\caption{This figure shows the knee of the ROC curves for two
simulations with $\rho=1/2$
and $\mbox{SNR}=1$. The left panel has $\mbox{SLR}=1/2$.
In this case the raw method beats SVA which
beats \mbox{EIGENSTRAT}.
The right panel has $\mbox{SLR}=1/4$ and SVA
beats EIGENSTRAT which beats the raw method.
In every case we simulated, the best
results are for an oracle that was given
the latent variables. The second best
was always for the proposed LEAPP method.
The relative performance
for SVA, EIGENSTRAT and the raw method
were different in other settings.}\label{figanROC}
\end{figure}

The ROC curves for two sets of conditions are shown in Figure
\ref{figanROC}. The best performance is always from the oracle. The
next best method is LEAPP. For the conditions in the left panel RAW is
next best followed by SVA and EIGENSTRAT. In the right panel SVA is
third, followed by EIGENSTRAT and then RAW.

Because the ROC curves from the simulations
have few if any crossings,
we can reasonably summarize each one by a single
number. We have used the area under the curve (AUC)
for a global comparison.
We also use a precision measure for the quality of\vadjust{\goodbreak}
the most highly ranked values. That measure
is the fraction of truly nonnull genes
among the highest ranking $H$ genes.
We use $H=50$.

When $\rho=0$, EIGENSTRAT, SVA and
LEAPP have almost equivalent performance.
For $\rho>0$,
the oracle always had the highest
AUC and LEAPP was always second.
The ordering among the other three methods
varied.
Sometimes \mbox{EIGENSTRAT} was the best
of those three, other times SVA
was the best of those three and other
times RAW was the best of those three.\looseness=-1

\begin{figure}

\includegraphics{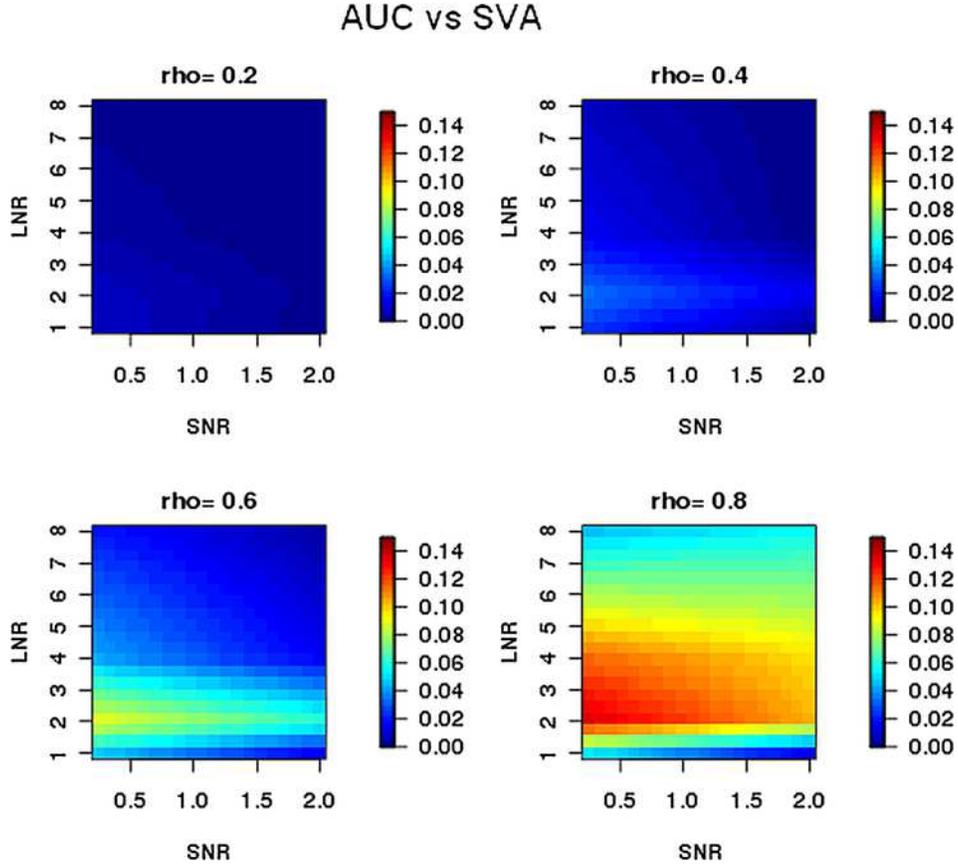}

\caption{This figure shows the improvement in AUC for LEAPP relative to
SVA. Here $\rho$ is the correlation between the primary and latent
variables. The signal to noise ratio and latent to noise ratio are
described in the text. The color scheme encodes
$(\mathrm{AUC}_{\mathrm{lea}}-\mathrm{AUC}_{\mathrm{sva}})/
\mathrm{AUC}_{\mathrm{sva}}$.}\label{figheatvssva}
\end{figure}

Figure~\ref{figheatvssva} shows a heatmap of the
improvement in AUC for LEAPP
versus SVA. The improvements are greatest
when $\rho$ is large. This is reasonable
because SVA is not designed to account
for correlation between the latent
and primary variables.
At each correlation level,
the greatest differences arise when
SNR is small and LNR is about $2$.

\begin{figure}

\includegraphics{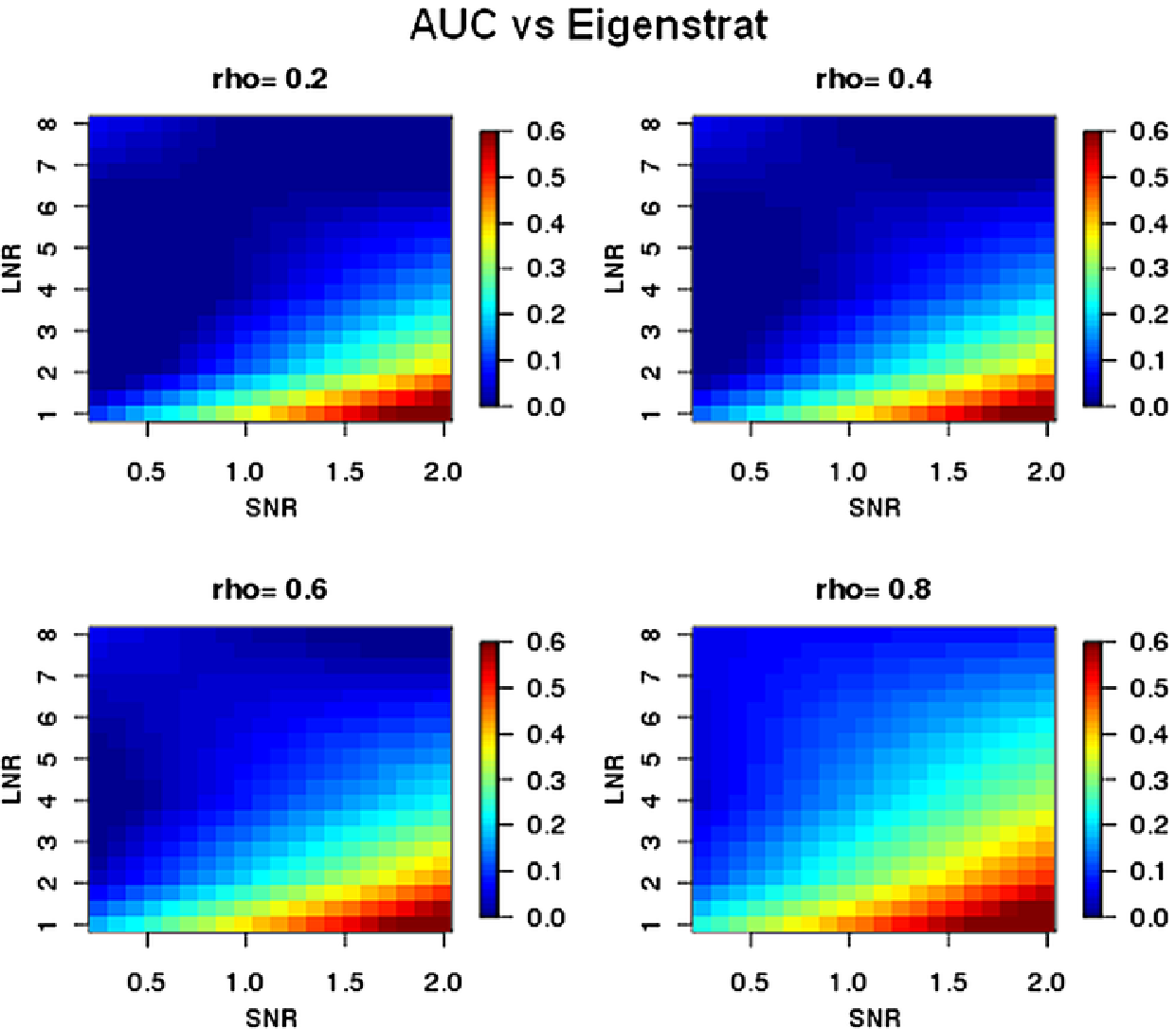}

\caption{This figure shows the improvement in AUC for LEAPP relative to
\mbox{EIGENSTRAT}. The simulation conditions are as described in
Figure \protect\ref{figheatvssva}. The color scheme encodes
$(\mathrm{AUC}_{\mathrm{rot}}-\mathrm{AUC}_{\mathrm{eig}})/
\mathrm{AUC}_{\mathrm{eig}}$.}\label{figheatvseig}
\end{figure}

Figure~\ref{figheatvseig} shows
the improvement in AUC for LEAPP
versus EIGENSTRAT.
The improvements are largest when
the primary effect is large.

The improvements versus SVA are smaller
than those versus EIGENSTRAT. To judge
the practical significance of the improvement,
we repeated some of these simulations
for SVA, increasing $n$ until SVA
achieved the same AUC that LEAPP
did. Sometimes SVA required only
$2$ more observations (one treatment
and one control) to match the AUC of
LEAPP. Sometimes it was unable
to match the AUC even given double the
sample size, that is, $n=120$ observations
instead of $n=60$. Not surprisingly, the
advantage of LEAPP is greatest when
the latent variable is most strongly
correlated with the primary.

\begin{table}
\def\arraystretch{0.9}
\caption{This table shows the number of samples required for SVA to
attain the same AUC that LEAPP attains with $n=60$ samples. For
example, with $\mbox{SNR}=2$ and $\mbox{LNR}=0.5$, and $\rho=0.25$,
SVA~requires $66$ samples or $10$\% more. The entries of $100$\% denote
settings where the increase needed was
\mbox{$\ge$}$100$\%}\label{tabextrasample}
\begin{tabular*}{\tablewidth}{@{\extracolsep{\fill}}ld{1.1}crcrrr@{}}
\hline
\multicolumn{2}{@{}l}{\textbf{Conditions}}
&\multicolumn{2}{c}{$\bolds{\rho=0.25}$}&
\multicolumn{2}{c}{$\bolds{\rho=0.5}$}&
\multicolumn{2}{c@{}}{$\bolds{\rho=0.75}$}\\[-4pt]
\multicolumn{2}{@{}l}{\hrulefill}
&\multicolumn{2}{c}{\hrulefill}&
\multicolumn{2}{c}{\hrulefill}&
\multicolumn{2}{c@{}}{\hrulefill}\\
\textbf{SNR} & \multicolumn{1}{c}{\textbf{LNR}}
& \multicolumn{1}{c}{$\bolds{n}$} & \multicolumn{1}{c}{\textbf{\%}}
& \multicolumn{1}{c}{$\bolds{n}$} & \textbf{\textbf{\%}}
& \multicolumn{1}{c}{$\bolds{n}$} & \multicolumn{1}{c@{}}{\textbf{\%}}\\
\hline
2&0.5&66 &10 &66 &10 &62 &3 \\
2&1 &68 &13 &92 &53 &120 &100 \\
2&2 &66 &10 &74 &23 &114 &90 \\
2&4 &62 &3 &66 &10 &88 &47 \\
2&8 &62 &3 &66 &10 &72 &20 \\
\hline
1&0.5 &64 &7 &64 &7 &62 &3 \\
1&1 &66 &10 &90 &50 &120 &100 \\
1&2 &64 &7 &76 &27 &120 &100 \\
1&4 &64 &7 &66 &10 &90 &50 \\
1&8 &62 &3 &66 &10 &76 &27 \\
\hline
0.5&0.5 &64 &7 &64 &7 &62 &3 \\
0.5&1 &66 &10 &84 &40 &120 &100 \\
0.5&2 &66 &10 &78 &30 &110 &83 \\
0.5&4 &66 &10 &68 &13 &88 &47 \\
0.5&8 &62 &3 &68 &13 &72 &20 \\
\hline
\end{tabular*}
\end{table}

Table~\ref{tabextrasample} shows a feature
of this problem that we also see in
the figures. The improvement over SVA
is quite small when $\mathrm{LNR}=0.5$.
A small enough latent effect becomes
undetectable, both methods suffer
and there is little difference.
Similarly, a very large latent
effect ($\mathrm{LNR}=8$) is easy to detect
by both methods. The largest differences
arise for medium sized latent effects.

\begin{figure}

\includegraphics{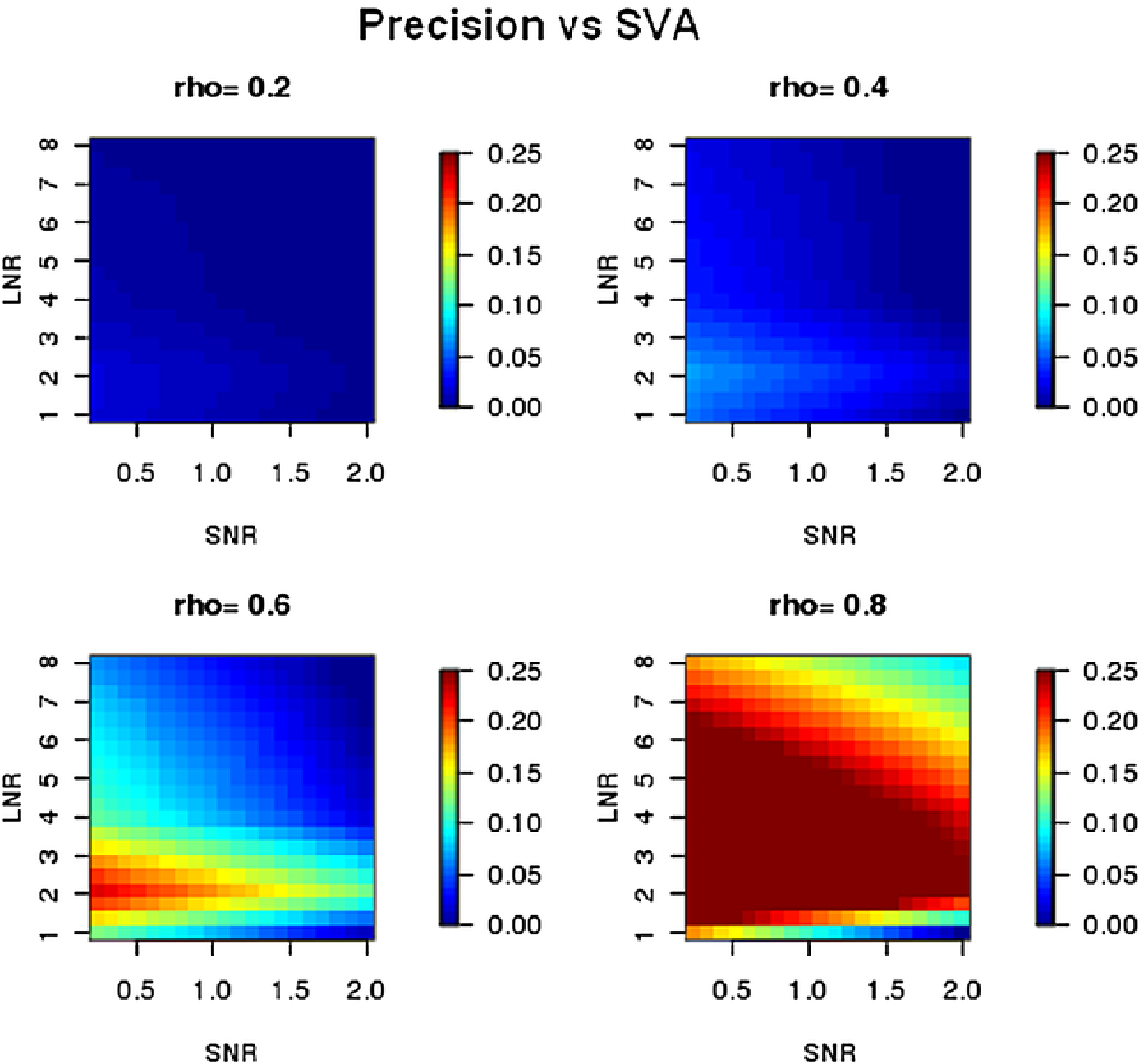}

\caption{This figure shows the improvement in precision for LEAPP
relative to SVA. Precision is the fraction of truly affected genes
among the top $H=50$ ranked genes. The simulation conditions are as
described in Figure \protect\ref{figheatvssva}. The color scheme
encodes $(\mathrm{PRE}_{\mathrm{lea}}-\mathrm{PRE}_{\mathrm{sva}})/
\mathrm{PRE}_{\mathrm{sva}}$.}\label{figheatvssvapre}
\end{figure}

\begin{figure}

\includegraphics{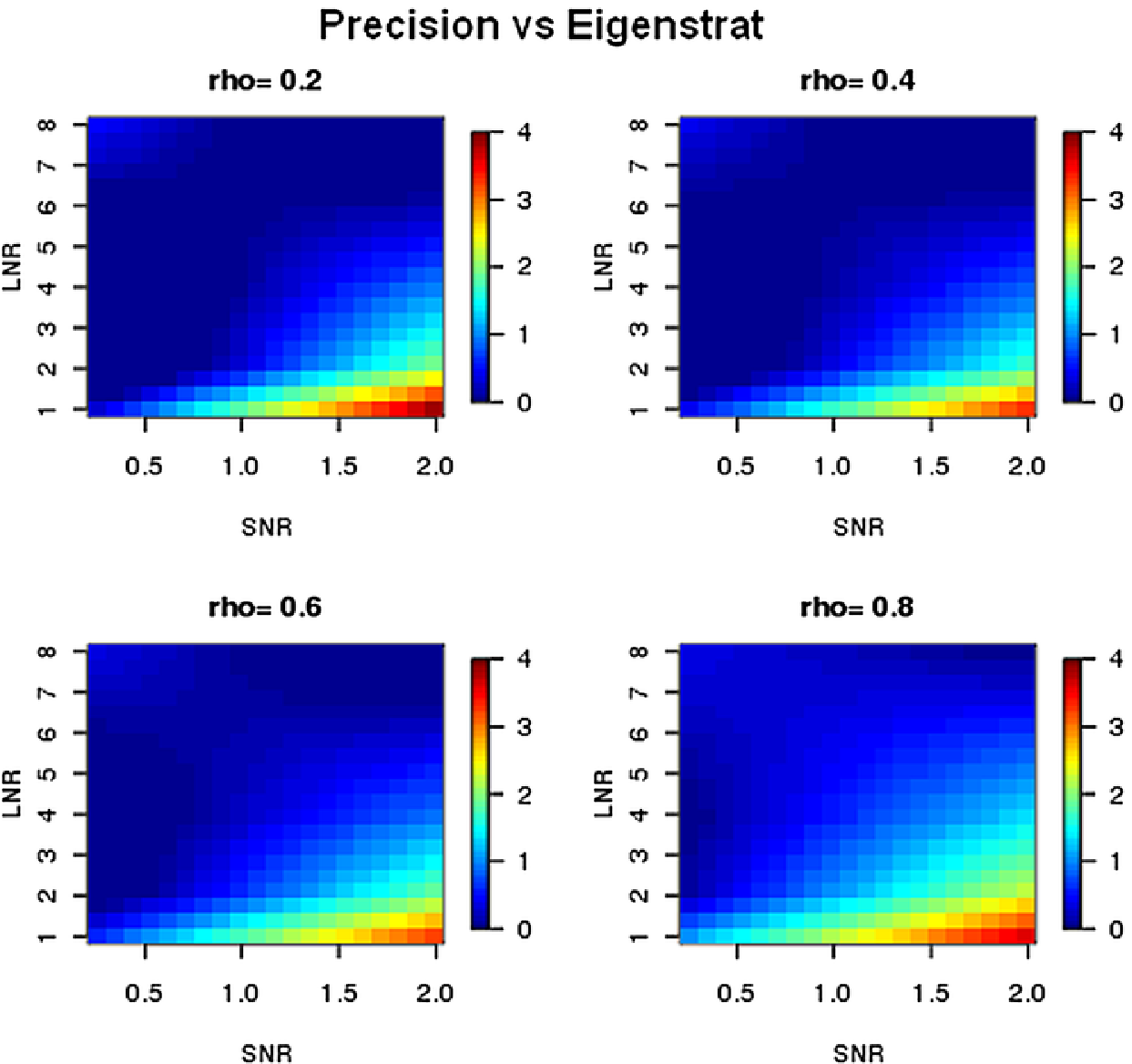}%
\vspace*{-3pt}
\caption{This figure shows the improvement in precision for LEAPP
relative to EIGENSTRAT. Precision is the fraction of truly affected
genes among the top $H=50$ ranked genes. The simulation conditions are
as described in Figure \protect\ref{figheatvssva}. The color scheme encodes
$(\mathrm{PRE}_{\mathrm{rot}}-\mathrm{PRE}_{\mathrm{eig}})/
\mathrm{PRE}_{\mathrm{eig}}$.}\label{figheatvseigpre}\vspace*{-3pt}
\end{figure}

High throughput methods are often used to identify candidates for
future follow-up investigation. In that case we value high precision
for the most highly ranked hypotheses. Figure~\ref{figheatvssvapre}
shows the improvement of LEAPP over SVA, as measured by precision.
Figure~\ref{figheatvseigpre} shows the improvement of LEAPP over
EIGENSTRAT, as measured by precision.

\section{AGEMAP data}\label{secagemap}

It is hard to find a real data set where the
true set of important genes is known. Even if
we are confident that a few genes are active, we
still cannot be sure that the others are really
inactive: the corresponding null hypotheses might be accepted, but
they are not proved.
We turn instead to the AGEMAP study [\citet{agemap}].

The AGEMAP study [\citet{agemap}]
investigated age-related
gene expression in mice.
Ten mice at each of four age groups were
investigated. From these $40$ mice,
samples were taken of $16$ different tissues,
resulting in $640$ microarray data sets.
A small number of those $640$ microarrays
were missing. From each microarray, $8932$
probes were sampled.
\citet{rotest} found that many of the
tissues in this data set exhibited strong latent variables.
Their approach assumed that the latent variables
were orthogonal to the treatment.

Our underlying assumption is that aging should
have partially though not totally consistent results from
tissue to tissue. According to \citet{kim2008}:
``Some aspects of aging only affect specific tissues;
examples include progressive weakness of muscle,
declining synaptic function in the brain, and decreased
filtration rate in the kidney. Other aspects of aging occur
in all cells regardless of their tissue type, such as the accumulation
of oxidative damage, and telomere shortening.''
\citet{muscle2006} found
some genetic pathways with common age regulation
in (human) kidney, brain and muscle.
\citet{grodetal2004} found common aging between
human kidney, cortex and medulla.
Some aspects of aging are also common
from species to species \citet{kim2007}.

A tendency for some common component
to aging should in turn produce
overlap in gene lists computed from
multiple tissues. Because age-related
genes are sparse, noisy estimation is
more likely to reduce overlap in gene
lists than to create it.

To illustrate this point, consider a
setting with $1000$ genes and two
tissues $A$ and $B$ with counts
\[
\bordermatrix{& A & \neg A
\cr
\hphantom{\neg }B & 10 & 10
\cr
\neg B & 10 & 970}.
\]
Here $10$ genes are truly age-related in both tissues,
$10$ are age-related in $A$ but not~$B$, and, finally,
$970$ genes are not age-related in either tissue.
Suppose now that statistical testing identifies each truly
age-related gene with power $0.6$ and
that each nonage-related gene has a false discovery
probability of $0.01$.
Using $\widehat A$ and $\widehat B$ to represent genes
identified as age-related, the expected counts
(for independent test statistics) are
in the following matrix:
\[
\bordermatrix{& \widehat A & \neg\widehat A
\cr
\hphantom{\neg}\widehat B & 3.817 & 17.983
\vspace*{4pt}\cr
\neg\widehat B &
17.983 & 960.217}.
\]
The effect of noisy gene identification is
severely biased toward reducing the apparent overlap.

For any two tissues, we can measure the
overlap between their sets of~highly ranked genes.
For two sets $A$ and $B$,
their resemblance [\citet{brod1997}] is
\[
\operatorname{res}(A,B) = \frac{ |A\cap B|}{ |A\cup B|},
\]
where $|\cdot|$ denotes cardinality.
Given two tissues and a significance level $\alpha$,
we can compute the resemblance of the
genes identified as age-related in the
tissues. Resemblance is then a function
of $\alpha$.
Plotting the numerator $|A\cap B|$ versus
the denominator $|A\cup B|$ as $\alpha$
increases, we obtain curves depicting
the strength of the overlap.\vadjust{\goodbreak}

In our setting with $16$
tissues there are ${16\choose2} = 120$
resemblances to consider. To keep the comparison
manageable as well as to pool information
from all tissues, we computed the following
quantities:
%
\begin{equation}
\label{eqIandU} I_\alpha= \sum_{1\le j <j'\le16}
\bigl|A_j^\alpha\cap A_{j'}^\alpha\bigr|
\quad\mbox{and}\quad U_\alpha= \Biggl|\bigcup_{j=1}^{16}
A_j^\alpha\Biggr|,
\end{equation}
where $A_j^\alpha$ is the set of statistically significant
genes at level $\alpha$ for tissue $j$.
We can think of $I_\alpha/U_\alpha$ as a pooled
resemblance. We would like to see large $I_\alpha$
at each given level of $U_\alpha$.

\begin{figure}

\includegraphics{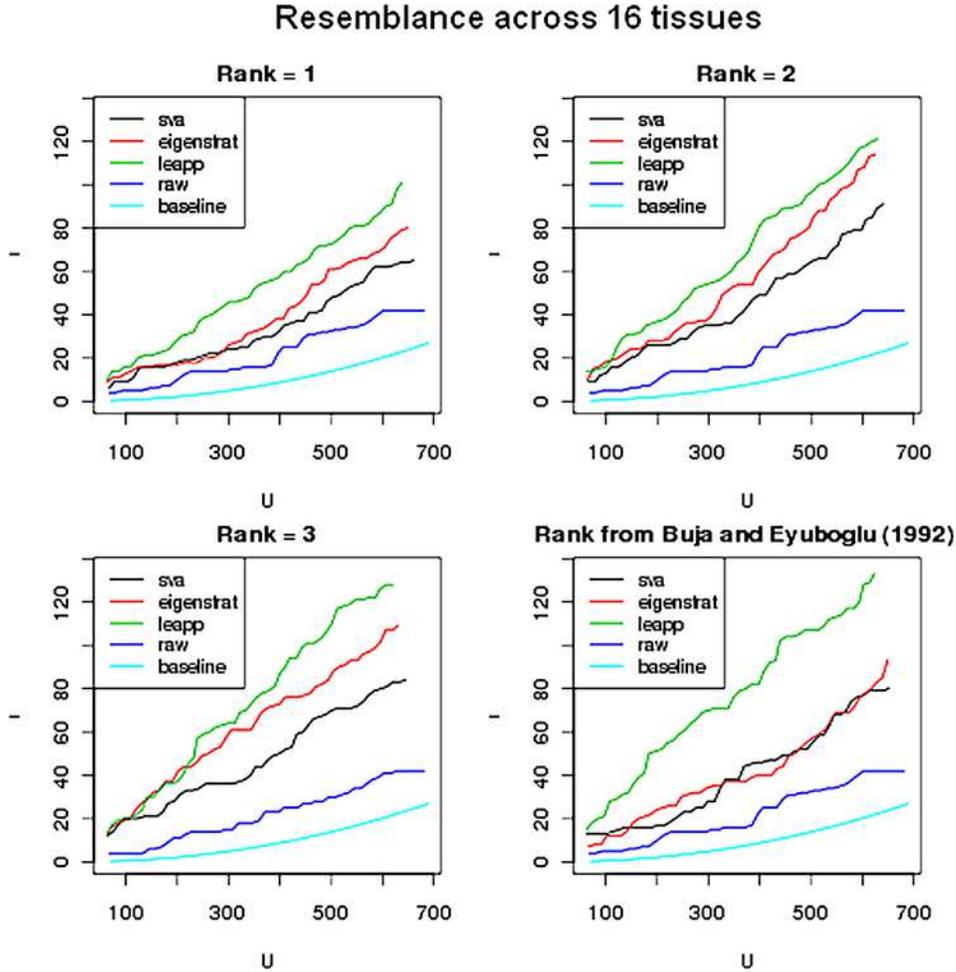}

\caption{This figure shows the resemblance among significant gene sets
from $16$ tissues in the AGEMAP study. We plot $I_\alpha$ versus
$U_\alpha$ [from equation (\protect\ref{eqIandU})], increasing
$\alpha$ from
$0$ until $U_\alpha=700$. The greatest self-consistency among lists is
from LEAPP. EIGENSTRAT is second best. The baseline curve is computed
assuming that the rankings for all $16$ tissues are mutually
independent.}\label{figresemb}
\end{figure}

Figure~\ref{figresemb} plots
$I_\alpha$ versus $U_\alpha$
for the methods we are comparing.
To make a precise comparison, we arranged
for each method that estimated latent
structure to employ the same estimate
for the rank of the latent component.
That rank is either $1$, $2$, $3$ or the
value chosen by the method of \citet{bujaeyob1992}.
At any rank LEAPP generates the
most self-consistent gene lists over almost
the entire range. EIGENSTRAT is usually second.
SVA beats a raw method that makes no adjustments.
LEAPP retains its strong performance
when the rank is chosen from the data while
the other two methods become poorer in that case.

Resemblance across tissues could also be high if there exists latent
variables strongly correlated with age which are repeated across
tissues. For example, consider a scenario where all tissues from young
mice are in one batch, and all tissues from elder mice are in a
different batch. If there are strong batch biases, then ``age-related''
genes would be reported by the raw method, and the same genes would be
ranked high across all tissues. However, note that raw performs the
worst of all methods in Figure~\ref{figresemb}, which gives some
reassurance that the high resemblance of the other methods is due to
successful removal of latent variables.

Given what we have learned from simulations,
the relative performance of EIGENSTRAT and SVA
gives us some insight into these data.
Since \mbox{EIGENSTRAT} has done well, it is more
likely that the signal is not very strong.
Since SVA has done poorly, it is more likely
that the latent variables in these data are
correlated with age. There is also the possibility
that they are correlated with sex (the covariate).
Our simulations did not include a covariate.

\section{Conclusions}\label{secconclusions}

High throughput testing has performance
that deteriorates in the presence of latent
variables. Latent variables that
are correlated with the treatment variable of interest
can severely alter the ordering of $p$-values.
Our LEAPP method separates the latent variable
from the treatment variable, making an adjustment
possible.

We have found in simulations that the adjustment
brings about a better ordering among hypotheses
than is available from either SVA or EIGENSTRAT.
The improvement over SVA is largest when
the latent variable is correlated with the
primary one. The improvement over EIGENSTRAT
is largest when the primary variable has a
large effect.

A referee asked about the case where the
coefficients of $\gamma$ for the
primary variable correlate over genes with the
per gene latent variable, $U$ in our notation.
We have not simulated such a case. It might
be very difficult for all methods or it might
be comparable to the case where $g$ correlates
with~$V$. It seems clear that
if $UV^\tran$ matches $\gamma g^\tran$ closely
enough, then it will be impossible to identify
relevant genes in this model.

In the simulations reported here
the data are drawn from the model under
which LEAPP was derived.
\citet{sun2011} also simulates the LEAPP, SVA
and \mbox{EIGENSTRAT} algorithms on the
model used by \citet{pric2006} to represent SNP
association studies. The SNPs themselves are
drawn from the Balding--Nichols model
[\citet{baldnich1995}]. Two scenarios were
considered. In both, the LEAPP ROC curve
placed above that for SVA which was above
that for EIGENSTRAT. All methods were
close when the relative risk for the causal
allele was $R=1.5$ while EIGENSTRAT
lagged behind for the case with $R=3$.

On the AGEMAP data we found
better consistency among tissues for significance
estimated by LEAPP than for either SVA
or EIGENSTRAT.

Some applications may have features measured
on the genes with per-sample covariates to be estimated
statistically. Such terms can be included in the
criss-cross regression framework but we have no
experience fitting them.

LEAPP produces $p$-values in addition to the
relative ordering of the genes. In this paper
we have only looked at the quality of the relative
ordering. In response to a reviewer's query
about calibration of $p$-values,
we created a QQ-plot of test statistics $T_i$
at (\ref{Ti}) on simulated data (not shown) and found
it very nearly linear. That simulated data was
pure noise, having no regression or latent
structure. For an investigation on real data,
Sun [(\citeyear{sun2011}), Chapter~4.5.2]
considered the breast cancer data
from \citet{hede2001}. She finds that the test
statistics produced by LEAPP have an empirical
null distribution from the R package locfdr
[\citet{efro2008}]
of $\dnorm(0.012,1.018^2)$ that closely matches
the nominal null distribution. That is what
we would expect to see if the nominal $p$-values
coming out of LEAPP had the
$U[0,1]$ distribution that they should have.
Corresponding empirical nulls are
$\dnorm(-0.01,1.55^2)$ for the RAW method,
$\dnorm(-0.009,1.425^2)$ for SVA and
$\dnorm(-0.093,1.199^2)$ for EIGENSTRAT.
Thus, in addition to a general improved ordering of
genes, this one example had
$p$-values that are better calibrated
in LEAPP than in SVA or EIGENSTRAT.

\begin{appendix}\label{app}
\section*{Appendix}

Here we give some properties
of our approach to testing many hypotheses
in the presence of latent variables.
We focus on a simpler version of the model
that is more tractable:
%
\begin{equation}
\label{eqsimpler} Y = \gamma g^\tran+ UV^\tran+\sigma E,
\end{equation}
where
$g\in\real^{n\times1}$ with $\Vert g\Vert=1$ as before,
$U\in\real^{N\times k}$ is nonrandom,
$V\in\real^{n\times k}$ has IID rows
with $\e(V^\tran V)= I_k$, known rank $k$
and $E\sim\dnorm(0,I_N\otimes I_n)$.
Compared to the full model (\ref{eqfull}),
equation (\ref{eqsimpler}) has no
covariate term $\beta X^\tran$, and has
constant variance $\Sigma= \sigma I_N$.

This simplification allows us to apply
results from the literature to our model.
It removes the Monte Carlo based
rank estimation step and the alternation
between estimating $\Sigma$ and using
the estimate $\widehat\Sigma$.
When $k=1$, the primary to latent correlation is
$\rho= g^\tran V/\sqrt{V^\tran V}$.

Our algorithm
requires the choice of a rotation matrix $O$ such that $Og = e_1$.
There are multiple possibilities for this matrix.
Our algorithm is invariant to the choice of $O$.
%
\begin{theorem}\label{thminvariant}
Let $Y$ follow the model (\ref{eqsimpler}).
Then our estimates of $U$ and
$\gamma$ do not depend on the rotation $O$ used
as long as $Og = e_1$.
\end{theorem}
\begin{pf}
See \citet{sun2011}.
\end{pf}

It is not hard to extend the proof
of Theorem~\ref{thminvariant}
to account for the $\beta X^\tran$ term.
The criss-cross regression begins
by computing $\widehat\beta$ from sums of
squares and cross-products. Those sums
of squares and cross-products are invariant
under the rotation.

The following theorem provides
a sufficient condition for our estimate $\widehat U$
to consistently estimate $U$.
We study the case where the data
are generated with $k=1$
and the model is also estimated using the
correct rank $k=1$.
Then as long as the latent factor $U$
is large enough compared
to the noise level, we will be able to detect and estimate $U$ fairly
well.
Our size measure $\Vert U\Vert^2_2(1-\rho^2)/n$
takes account of the correlation. With a higher
$\rho$, more of the latent factor is removed
from $Y^{(\ell)}$.

We measure error by the cosine
$\Phi(\widehat U,U)= \widehat U^\tran U/(\Vert\widehat U\Vert_2 \Vert U\Vert_2)$
of the angle between $\widehat U$ and $U$.
The estimate $\widehat U$ is determined only
up to sign. Replacing $\widehat U$ by $-\widehat U$
causes a change from
$\widehat V$ to $-\widehat V$ and leaves the model unchanged.
We only need
$\max(\Phi(\widehat U,U),\Phi(-\widehat U,U))
=|\Phi(\widehat U,U)|\to1$ for consistency.
%
\begin{theorem}\label{thmUerr}
Let $Y$ follow the model (\ref{eqsimpler}) with $k=1$
and ${\Vert U\Vert_2^2 (1-\rho^2)}/\allowbreak{n} \rightarrow
\infty$ and ${N(n)}/{n} \rightarrow c \in(0,\infty)$ as $n\to\infty$.
Let $\widehat{U}$ be our estimator for $U$ using
$k=1$.
Then $|\Phi(\widehat{U},U)|
\rightarrow1$ as $n \rightarrow\infty$ with probability $1$.
\end{theorem}
\begin{pf}
See \citet{sun2011}.
\end{pf}

Next we give conditions for the final
step of LEAPP to accurately estimate~$\gamma$,
that is, for $\Vert\widehat\gamma-\gamma\Vert_2$
to be small.
To do this, we combine methods
used in random matrix theory from \citet{bai2003}
with methods used in
compressed sensing in \citet{candrand2006}.

In our simulations we found little difference
between robust and nonrobust versions
of the $\Theta$--IPOD algorithm. This is not
surprising, since our simulations did not
place nonzero $\gamma_i$ preferentially
at high leverage points (extreme $U_{i1}$). For our analysis
we replace the robust $\Theta$--IPOD algorithm
by the Dantzig selector for which
strong results are available.\vadjust{\goodbreak}

Our algorithm was designed assuming that
the primary variable $g$ is not too
strongly correlated with the
latent variable $V$.
In our analysis we also impose a separation
between the effects $\gamma$ and
the latent quantity $U$. Specifically,
we assume that $\gamma$ is sparse
and that $U$ is not.

The vector $x$ is $s$-sparse if it has
at most $s$ nonzero components.
Following \citet{candrand2006},
we define the sequences $a_s(A)$ and $b_s(A)$ as the largest and smallest
numbers, respectively, such that
\[
a_s(A)\Vert x\Vert_2 \leq\Vert Ax\Vert_2
\leq b_s(A)\Vert x\Vert_2
\]
holds for all $s$-sparse $x$.
%
\begin{theorem}\label{thmgammaerr}
Suppose that $Y$ follows the
model (\ref{eqsimpler}) with $k=1$,
a~fixed correlation $\rho\in(-1,1)$
between $g$ and $V$,
and an $s$-sparse vector $\gamma$.
Assume that ${N}/{n} \rightarrow c \in(0,\infty)$, $V^\tran V
\stackrel{p}{\rightarrow} 1$, and $(Nn)^{-1}\Vert U\Vert_2^2
\rightarrow
\sigma^2_u > 0$ hold as $n \rightarrow\infty$.
Let our estimated $U$ be $\widehat{U}$
and set $U^\star=\widehat{U}/\Vert\widehat U\Vert_2$.
Writing
$|U_{(1)}^\star| \geq|U_{(2)}^\star| \geq\cdots\geq|U_{(N)}^\star|$
for the ordered components of $U^\star$, assume
that there is a constant $0 < B < 1$ such that
\[
\sum_{i=1}^{2s} \bigl(U_{(i)}^\star
\bigr)^2 + \frac12\sum_{i=1}^{3s}
\bigl(U_{(i)}^\star\bigr)^2 \leq B.
\]
Then the Dantzig estimator $\widehat{\gamma}$, which minimizes
\[
\Vert\widehat\gamma\Vert_1 \quad\mbox{subject to}\quad \bigl\Vert\bigl(I -
U^\star U^{\star\tran}\bigr) \bigl(Y_1^{(r)} -
\widehat\gamma\bigr) \bigr\Vert_\infty\leq\sigma\sqrt{\log N}
\]
satisfies
\[
\Vert\widehat{\gamma} - \gamma\Vert_2^2 \leq
\frac{ 16 \sigma^2 s\log(N)}{(1-\rho^2) (1-B)^2}.
\]
\end{theorem}
\begin{pf}
See \citet{sun2011}.
\end{pf}
\end{appendix}



\printaddresses


\begin{thebibliography}{36}

\bibitem[\protect\citeauthoryear{Allen and Tibshirani}{2010}]{alletibs2010tr}
\begin{btechreport}[author]
\bauthor{\bsnm{Allen},~\bfnm{G.~I.}\binits{G.~I.}} \AND
  \bauthor{\bsnm{Tibshirani},~\bfnm{R.~J.}\binits{R.~J.}}
(\byear{2010}).
\btitle{Inference with transposable data: Modeling the effects of row and
  column correlations}.
\btype{Technical report},
\blocation{Stanford Univ., Dept. Statistics}.
\bptok{imsref}%
\end{btechreport}
\endbibitem

\bibitem[\protect\citeauthoryear{Bai}{2003}]{bai2003}
\begin{barticle}[mr]
\bauthor{\bsnm{Bai},~\bfnm{Jushan}\binits{J.}}
(\byear{2003}).
\btitle{Inferential theory for factor models of large dimensions}.
\bjournal{Econometrica}
\bvolume{71}
\bpages{135--171}.
\bid{doi={10.1111/1468-0262.00392}, issn={0012-9682}, mr={1956857}}
\bptok{imsref}%
\end{barticle}
\endbibitem

\bibitem[\protect\citeauthoryear{Balding and Nicols}{1995}]{baldnich1995}
\begin{barticle}[author]
\bauthor{\bsnm{Balding},~\bfnm{D.}\binits{D.}} \AND
  \bauthor{\bsnm{Nicols},~\bfnm{R.}\binits{R.}}
(\byear{1995}).
\btitle{A method for quantifying differentiation between populations at
  multi-allelic loci and its implications for investigating identity and
  paternity}.
\bjournal{Genetica}
\bvolume{96}
\bpages{3--12}.
\bptok{imsref}%
\end{barticle}
\endbibitem

\bibitem[\protect\citeauthoryear{Broder}{1997}]{brod1997}
\begin{binproceedings}[author]
\bauthor{\bsnm{Broder},~\bfnm{A.~Z.}\binits{A.~Z.}}
(\byear{1997}).
\btitle{On the resemblance and containment of documents}.
In \bbooktitle{Compression and Complexity of Sequences 1997. Proceedings}
\bpages{21--29}.
\bpublisher{IEEE Comput. Soc.}, \blocation{Los Alamitos}.
\bptok{imsref}%
\end{binproceedings}
\endbibitem

\bibitem[\protect\citeauthoryear{Buja and Eyuboglu}{1992}]{bujaeyob1992}
\begin{barticle}[author]
\bauthor{\bsnm{Buja},~\bfnm{A.}\binits{A.}} \AND
  \bauthor{\bsnm{Eyuboglu},~\bfnm{N.}\binits{N.}}
(\byear{1992}).
\btitle{Remarks on parallel analysis}.
\bjournal{Multivariate Behavioral Research}
\bvolume{27}
\bpages{509--540}.
\bptok{imsref}%
\end{barticle}
\endbibitem

\bibitem[\protect\citeauthoryear{Cand{\`e}s and Randall}{2006}]{candrand2006}
\begin{barticle}[mr]
\bauthor{\bsnm{Cand{\`e}s},~\bfnm{Emmanuel~J.}\binits{E.~J.}} \AND
  \bauthor{\bsnm{Randall},~\bfnm{Paige~A.}\binits{P.~A.}}
(\byear{2006}).
\btitle{Highly robust error correction by convex programming}.
\bjournal{IEEE Trans. Inform. Theory}
\bvolume{54}
\bpages{2829--2840}.
\bid{doi={10.1109/TIT.2008.924688}, issn={0018-9448}, mr={2450835}}
\bptnote{check year}%
\bptok{imsref}%
\end{barticle}\vadjust{\goodbreak}
\endbibitem

\bibitem[\protect\citeauthoryear{Carvalho et~al.}{2008}]{carvetal2008}
\begin{barticle}[mr]
\bauthor{\bsnm{Carvalho},~\bfnm{Carlos~M.}\binits{C.~M.}},
  \bauthor{\bsnm{Chang},~\bfnm{Jeffrey}\binits{J.}},
  \bauthor{\bsnm{Lucas},~\bfnm{Joseph~E.}\binits{J.~E.}},
  \bauthor{\bsnm{Nevins},~\bfnm{Joseph~R.}\binits{J.~R.}},
  \bauthor{\bsnm{Wang},~\bfnm{Quanli}\binits{Q.}} \AND
  \bauthor{\bsnm{West},~\bfnm{Mike}\binits{M.}}
(\byear{2008}).
\btitle{High-dimensional sparse factor modeling: Applications in gene
  expression genomics}.
\bjournal{J. Amer. Statist. Assoc.}
\bvolume{103}
\bpages{1438--1456}.
\bid{doi={10.1198/016214508000000869}, issn={0162-1459}, mr={2655722}}
\bptok{imsref}%
\end{barticle}
\endbibitem

\bibitem[\protect\citeauthoryear{Chen and Chen}{2008}]{chenchen2008}
\begin{barticle}[author]
\bauthor{\bsnm{Chen},~\bfnm{J.}\binits{J.}} \AND
  \bauthor{\bsnm{Chen},~\bfnm{Z.}\binits{Z.}}
(\byear{2008}).
\btitle{Extended {Bayesian} information criterion}.
\bjournal{Biometrika}
\bvolume{94}
\bpages{759--771}.
\bptok{imsref}%
\end{barticle}
\endbibitem

\bibitem[\protect\citeauthoryear{Diskin et~al.}{2008}]{disketal2008}
\begin{barticle}[pbm]
\bauthor{\bsnm{Diskin},~\bfnm{Sharon~J.}\binits{S.~J.}},
  \bauthor{\bsnm{Li},~\bfnm{Mingyao}\binits{M.}},
  \bauthor{\bsnm{Hou},~\bfnm{Cuiping}\binits{C.}},
  \bauthor{\bsnm{Yang},~\bfnm{Shuzhang}\binits{S.}},
  \bauthor{\bsnm{Glessner},~\bfnm{Joseph}\binits{J.}},
  \bauthor{\bsnm{Hakonarson},~\bfnm{Hakon}\binits{H.}},
  \bauthor{\bsnm{Bucan},~\bfnm{Maja}\binits{M.}},
  \bauthor{\bsnm{Maris},~\bfnm{John~M.}\binits{J.~M.}} \AND
  \bauthor{\bsnm{Wang},~\bfnm{Kai}\binits{K.}}
(\byear{2008}).
\btitle{Adjustment of genomic waves in signal intensities from whole-genome SNP
  genotyping platforms}.
\bjournal{Nucleic Acids Res.}
\bvolume{36}
\bpages{e126}.
\bid{doi={10.1093/nar/gkn556}, issn={1362-4962}, pii={gkn556}, pmcid={2577347},
  pmid={18784189}}
\bptok{imsref}%
\end{barticle}
\endbibitem

\bibitem[\protect\citeauthoryear{Dudoit and van~der Laan}{2008}]{dudovand2008}
\begin{bbook}[author]
\bauthor{\bsnm{Dudoit},~\bfnm{S.}\binits{S.}} \AND \bauthor{\bparticle{van~der}
  \bsnm{Laan},~\bfnm{M.~J.}\binits{M.~J.}}
(\byear{2008}).
\btitle{Multiple Testing Procedures with Applications to Genetics}.
\bpublisher{Springer}, \blocation{New York}.
\bptok{imsref}%
\end{bbook}
\endbibitem

\bibitem[\protect\citeauthoryear{Efron}{2007}]{efro2007}
\begin{barticle}[mr]
\bauthor{\bsnm{Efron},~\bfnm{Bradley}\binits{B.}}
(\byear{2007}).
\btitle{Size, power and false discovery rates}.
\bjournal{Ann. Statist.}
\bvolume{35}
\bpages{1351--1377}.
\bid{doi={10.1214/009053606000001460}, issn={0090-5364}, mr={2351089}}
\bptok{imsref}%
\end{barticle}
\endbibitem

\bibitem[\protect\citeauthoryear{Efron}{2008}]{efro2008}
\begin{barticle}[mr]
\bauthor{\bsnm{Efron},~\bfnm{Bradley}\binits{B.}}
(\byear{2008}).
\btitle{Microarrays, empirical {B}ayes and the two-groups model}.
\bjournal{Statist. Sci.}
\bvolume{23}
\bpages{1--22}.
\bid{doi={10.1214/07-STS236}, issn={0883-4237}, mr={2431866}}
\bptok{imsref}%
\end{barticle}
\endbibitem

\bibitem[\protect\citeauthoryear{Efron}{2010}]{efro2010}
\begin{bbook}[mr]
\bauthor{\bsnm{Efron},~\bfnm{Bradley}\binits{B.}}
(\byear{2010}).
\btitle{Large-Scale Inference:
Empirical Bayes Methods for Estimation, Testing, and Prediction}.
\bseries{Institute of Mathematical Statistics Monographs}
\bvolume{1}.
\bpublisher{Cambridge Univ. Press}, \blocation{Cambridge}.
\bid{mr={2724758}}
\bptok{imsref}%
\end{bbook}
\endbibitem

\bibitem[\protect\citeauthoryear{Friguet, Kloareg and
  Causeur}{2009}]{frigkloacasu2009}
\begin{barticle}[mr]
\bauthor{\bsnm{Friguet},~\bfnm{Chlo{\'e}}\binits{C.}},
  \bauthor{\bsnm{Kloareg},~\bfnm{Maela}\binits{M.}} \AND
  \bauthor{\bsnm{Causeur},~\bfnm{David}\binits{D.}}
(\byear{2009}).
\btitle{A factor model approach to multiple testing under dependence}.
\bjournal{J. Amer. Statist. Assoc.}
\bvolume{104}
\bpages{1406--1415}.
\bid{doi={10.1198/jasa.2009.tm08332}, issn={0162-1459}, mr={2750571}}
\bptok{imsref}%
\end{barticle}
\endbibitem

\bibitem[\protect\citeauthoryear{Gabriel and Zamir}{1979}]{gabrzami1979}
\begin{barticle}[author]
\bauthor{\bsnm{Gabriel},~\bfnm{K.~R.}\binits{K.~R.}} \AND
  \bauthor{\bsnm{Zamir},~\bfnm{S.}\binits{S.}}
(\byear{1979}).
\btitle{Lower rank approximation of matrices by least squares with any choice
  of weights}.
\bjournal{Technometrics}
\bvolume{21}
\bpages{489--498}.
\bptok{imsref}%
\end{barticle}
\endbibitem

\bibitem[\protect\citeauthoryear{Hedenfalk}{2001}]{hede2001}
\begin{barticle}[pbm]
\bauthor{\bsnm{Hedenfalk},~\bfnm{I.}\binits{I.}}
(\byear{2001}).
\btitle{Gene-expression profiles in hereditary breast cancer}.
\bjournal{N. Engl. J.~Med.}
\bvolume{344}
\bpages{539--548}.
\bptok{imsref}%
\end{barticle}
\endbibitem

\bibitem[\protect\citeauthoryear{Johnstone}{2001}]{john2001}
\begin{barticle}[mr]
\bauthor{\bsnm{Johnstone},~\bfnm{Iain~M.}\binits{I.~M.}}
(\byear{2001}).
\btitle{On the distribution of the largest eigenvalue in principal components
  analysis}.
\bjournal{Ann. Statist.}
\bvolume{29}
\bpages{295--327}.
\bid{doi={10.1214/aos/1009210544}, issn={0090-5364}, mr={1863961}}
\bptok{imsref}%
\end{barticle}
\endbibitem

\bibitem[\protect\citeauthoryear{Kim}{2007}]{kim2007}
\begin{barticle}[pbm]
\bauthor{\bsnm{Kim},~\bfnm{Stuart~K.}\binits{S.~K.}}
(\byear{2007}).
\btitle{Common aging pathways in worms, flies, mice and humans}.
\bjournal{J. Exp. Biol.}
\bvolume{210}
\bpages{1607--1612}.
\bid{doi={10.1242/jeb.004887}, issn={0022-0949}, pii={210/9/1607},
  pmid={17449826}}
\bptok{imsref}%
\end{barticle}
\endbibitem

\bibitem[\protect\citeauthoryear{Kim}{2008}]{kim2008}
\begin{bincollection}[author]
\bauthor{\bsnm{Kim},~\bfnm{S.~K.}\binits{S.~K.}}
(\byear{2008}).
\btitle{Genome-wide views of aging gene networks}.
In \bbooktitle{Molecular Biology of Aging}
\bpages{215--235}.
\bpublisher{Cold Spring Harbor Laboratory Press}, \blocation{Cold Spring
  Harbor, NY}.
\bptok{imsref}%
\end{bincollection}
\endbibitem

\bibitem[\protect\citeauthoryear{Leek and Storey}{2008}]{leekstor2008}
\begin{barticle}[author]
\bauthor{\bsnm{Leek},~\bfnm{J.~T.}\binits{J.~T.}} \AND
  \bauthor{\bsnm{Storey},~\bfnm{J.~D.}\binits{J.~D.}}
(\byear{2008}).
\btitle{A general framework for multiple testing dependence}.
\bjournal{Proc. Natl. Acad. Sci. USA}
\bvolume{105}
\bpages{18718--18723}.
\bptok{imsref}%
\end{barticle}
\endbibitem

\bibitem[\protect\citeauthoryear{Leek
  et~al.}{2010}]{leekschabravsimclongjohngemabaggiriz2010}
\begin{barticle}[author]
\bauthor{\bsnm{Leek},~\bfnm{J.~T.}\binits{J.~T.}},
  \bauthor{\bsnm{Scharpf},~\bfnm{R.~B.}\binits{R.~B.}},
  \bauthor{\bsnm{Corrada-Bravo},~\bfnm{H.}\binits{H.}},
  \bauthor{\bsnm{Simcha},~\bfnm{D.}\binits{D.}},
  \bauthor{\bsnm{Langmead},~\bfnm{B.}\binits{B.}},
  \bauthor{\bsnm{Johnson},~\bfnm{W.~E.}\binits{W.~E.}},
  \bauthor{\bsnm{Geman},~\bfnm{D.}\binits{D.}},
  \bauthor{\bsnm{Baggerley},~\bfnm{K.}\binits{K.}} \AND
  \bauthor{\bsnm{Irizarry},~\bfnm{R.~A.}\binits{R.~A.}}
(\byear{2010}).
\btitle{Tackling the widespread and critical impact of batch effects in
  high-throughput data}.
\bjournal{Nature Reviews Genetics}
\bvolume{11}
\bpages{733--739}.
\bptok{imsref}%
\end{barticle}
\endbibitem

\bibitem[\protect\citeauthoryear{Lucas, Kung and Chi}{2010}]{lucakungchi2010}
\begin{barticle}[author]
\bauthor{\bsnm{Lucas},~\bfnm{J.~E.}\binits{J.~E.}},
  \bauthor{\bsnm{Kung},~\bfnm{H.~N.}\binits{H.~N.}} \AND
  \bauthor{\bsnm{Chi},~\bfnm{J.~T.~A.}\binits{J.~T.~A.}}
(\byear{2010}).
\btitle{Latent factor analysis to discover pathway-associated putative
  segmental aneuploidies in human cancers}.
\bjournal{PLoS Comput. Biol.}
\bvolume{6}
\bpages{e100920:1--15}.
\bptok{imsref}%
\end{barticle}
\endbibitem

\bibitem[\protect\citeauthoryear{Olshen et~al.}{2004}]{olshvenkluciwigl2004}
\begin{barticle}[pbm]
\bauthor{\bsnm{Olshen},~\bfnm{Adam~B.}\binits{A.~B.}},
  \bauthor{\bsnm{Venkatraman},~\bfnm{E.~S.}\binits{E.~S.}},
  \bauthor{\bsnm{Lucito},~\bfnm{Robert}\binits{R.}} \AND
  \bauthor{\bsnm{Wigler},~\bfnm{Michael}\binits{M.}}
(\byear{2004}).
\btitle{Circular binary segmentation for the analysis of array-based DNA copy
  number data}.
\bjournal{Biostatistics}
\bvolume{5}
\bpages{557--572}.
\bid{doi={10.1093/biostatistics/kxh008}, issn={1465-4644}, pii={5/4/557},
  pmid={15475419}}
\bptok{imsref}%
\end{barticle}
\endbibitem

\bibitem[\protect\citeauthoryear{Owen and Perry}{2009}]{bicros}
\begin{barticle}[author]
\bauthor{\bsnm{Owen},~\bfnm{A.~B.}\binits{A.~B.}} \AND
  \bauthor{\bsnm{Perry},~\bfnm{P.~O.}\binits{P.~O.}}
(\byear{2009}).
\btitle{Bi-cross-validation of the {SVD} and the non-negative matrix
  factorization}.
\bjournal{Ann. Appl. Stat.}
\bvolume{3}
\bpages{564--594}.
\bptok{imsref}%
\end{barticle}
\endbibitem

\bibitem[\protect\citeauthoryear{Patterson, Price and
  Reich}{2006}]{pattpricreic2006}
\begin{barticle}[author]
\bauthor{\bsnm{Patterson},~\bfnm{N.~J.}\binits{N.~J.}},
  \bauthor{\bsnm{Price},~\bfnm{A.~L.}\binits{A.~L.}} \AND
  \bauthor{\bsnm{Reich},~\bfnm{D.}\binits{D.}}
(\byear{2006}).
\btitle{Population structure and eigenanalysis}.
\bjournal{PLoS Genetics}
\bvolume{2}
\bpages{2074--2093}.
\bptok{imsref}%
\end{barticle}
\endbibitem

\bibitem[\protect\citeauthoryear{Perry}{2009}]{perr2009}
\begin{bphdthesis}[author]
\bauthor{\bsnm{Perry},~\bfnm{P.~O.}\binits{P.~O.}}
(\byear{2009}).
\btitle{Cross-validation for unsupervised learning}.
\btype{Ph.D. thesis},
\blocation{Stanford Univ}.
\bptok{imsref}%
\end{bphdthesis}
\endbibitem

\bibitem[\protect\citeauthoryear{Perry and Owen}{2010}]{rotest}
\begin{barticle}[mr]
\bauthor{\bsnm{Perry},~\bfnm{Patrick~O.}\binits{P.~O.}} \AND
  \bauthor{\bsnm{Owen},~\bfnm{Art~B.}\binits{A.~B.}}
(\byear{2010}).
\btitle{A rotation test to verify latent structure}.
\bjournal{J.~Mach. Learn. Res.}
\bvolume{11}
\bpages{603--624}.
\bid{issn={1532-4435}, mr={2600622}}
\bptok{imsref}%
\end{barticle}
\endbibitem

\bibitem[\protect\citeauthoryear{Price et~al.}{2006}]{pric2006}
\begin{barticle}[author]
\bauthor{\bsnm{Price},~\bfnm{A.~L.}\binits{A.~L.}},
  \bauthor{\bsnm{Patterson},~\bfnm{N.~J.}\binits{N.~J.}},
  \bauthor{\bsnm{Plengt},~\bfnm{R.~M.}\binits{R.~M.}},
  \bauthor{\bsnm{Weinblatt},~\bfnm{M.~E.}\binits{M.~E.}},
  \bauthor{\bsnm{Shadick},~\bfnm{N.~A.}\binits{N.~A.}} \AND
  \bauthor{\bsnm{Reich},~\bfnm{D.}\binits{D.}}
(\byear{2006}).
\btitle{Principal components ananysis corrects for stratification in
  genome-wide association studies}.
\bjournal{Nature Genetics}
\bvolume{38}
\bpages{904--909}.
\bptok{imsref}%
\end{barticle}
\endbibitem

\bibitem[\protect\citeauthoryear{Rodwell et~al.}{2004}]{grodetal2004}
\begin{barticle}[author]
\bauthor{\bsnm{Rodwell},~\bfnm{G.}\binits{G.}},
  \bauthor{\bsnm{Sonu},~\bfnm{R.}\binits{R.}},
  \bauthor{\bsnm{Zahn},~\bfnm{J.~M.}\binits{J.~M.}},
  \bauthor{\bsnm{Lund},~\bfnm{J.}\binits{J.}},
  \bauthor{\bsnm{Wilhelmy},~\bfnm{J.}\binits{J.}},
  \bauthor{\bsnm{Wang},~\bfnm{L.}\binits{L.}},
  \bauthor{\bsnm{Xiao},~\bfnm{W.}\binits{W.}},
  \bauthor{\bsnm{Mindrinos},~\bfnm{M.}\binits{M.}},
  \bauthor{\bsnm{Crane},~\bfnm{E.}\binits{E.}},
  \bauthor{\bsnm{Segal},~\bfnm{E.}\binits{E.}},
  \bauthor{\bsnm{Myers},~\bfnm{B.}\binits{B.}},
  \bauthor{\bsnm{Davis},~\bfnm{R.}\binits{R.}},
  \bauthor{\bsnm{Higgins},~\bfnm{J.}\binits{J.}},
  \bauthor{\bsnm{Owen},~\bfnm{A.~B.}\binits{A.~B.}} \AND
  \bauthor{\bsnm{Kim},~\bfnm{S.~K.}\binits{S.~K.}}
(\byear{2004}).
\btitle{A transcriptional profile of aging in the human kidney}.
\bjournal{PLoS Biology}
\bvolume{2}
\bpages{2191--2201}.
\bptok{imsref}%
\end{barticle}
\endbibitem

\bibitem[\protect\citeauthoryear{She and Owen}{2011}]{sheowen2011}
\begin{barticle}[author]
\bauthor{\bsnm{She},~\bfnm{Y.}\binits{Y.}} \AND
  \bauthor{\bsnm{Owen},~\bfnm{A.~B.}\binits{A.~B.}}
(\byear{2011}).
\btitle{Outlier identification using nonconvex penalized regression}.
\bjournal{J.~Amer. Statist. Assoc.}
\bvolume{106}
\bpages{626--639}.
\bptok{imsref}%
\end{barticle}
\endbibitem

\bibitem[\protect\citeauthoryear{Storey, Akey and
  Kruglyak}{2005}]{storakeykrug2005}
\begin{barticle}[author]
\bauthor{\bsnm{Storey},~\bfnm{J.~D.}\binits{J.~D.}},
  \bauthor{\bsnm{Akey},~\bfnm{J.~M.}\binits{J.~M.}} \AND
  \bauthor{\bsnm{Kruglyak},~\bfnm{L.}\binits{L.}}
(\byear{2005}).
\btitle{Multiple locus linkage analysis of genomewide expression in yeast}.
\bjournal{PLoS Biology}
\bvolume{3}
\bpages{1380--1390}.
\bptok{imsref}%
\end{barticle}
\endbibitem

\bibitem[\protect\citeauthoryear{Sun}{2011}]{sun2011}
\begin{bphdthesis}[author]
\bauthor{\bsnm{Sun},~\bfnm{Y.}\binits{Y.}}
(\byear{2011}).
\btitle{On latent systemic effects in multiple hypotheses}.
\btype{Ph.D. thesis},
\blocation{Stanford Univ}.
\bptok{imsref}%
\end{bphdthesis}
\endbibitem

\bibitem[\protect\citeauthoryear{Tracy and Widom}{1994}]{tracwido1994}
\begin{barticle}[mr]
\bauthor{\bsnm{Tracy},~\bfnm{Craig~A.}\binits{C.~A.}} \AND
  \bauthor{\bsnm{Widom},~\bfnm{Harold}\binits{H.}}
(\byear{1994}).
\btitle{Level-spacing distributions and the {A}iry kernel}.
\bjournal{Comm. Math. Phys.}
\bvolume{159}
\bpages{151--174}.
\bid{issn={0010-3616}, mr={1257246}}
\bptok{imsref}%
\end{barticle}
\endbibitem

\bibitem[\protect\citeauthoryear{Zahn et~al.}{2006}]{muscle2006}
\begin{barticle}[author]
\bauthor{\bsnm{Zahn},~\bfnm{J.~M.}\binits{J.~M.}},
  \bauthor{\bsnm{Sonu},~\bfnm{R.}\binits{R.}},
  \bauthor{\bsnm{Vogel},~\bfnm{H.}\binits{H.}},
  \bauthor{\bsnm{Crane},~\bfnm{E.}\binits{E.}},
  \bauthor{\bsnm{Mazan-Mamczarz},~\bfnm{K.}\binits{K.}},
  \bauthor{\bsnm{Rabkin},~\bfnm{R.}\binits{R.}},
  \bauthor{\bsnm{Davis},~\bfnm{R.~W.}\binits{R.~W.}},
  \bauthor{\bsnm{Becker},~\bfnm{K.~G.}\binits{K.~G.}},
  \bauthor{\bsnm{Owen},~\bfnm{A.~B.}\binits{A.~B.}} \AND
  \bauthor{\bsnm{Kim},~\bfnm{S.~K.}\binits{S.~K.}}
(\byear{2006}).
\btitle{Transcriptional profiling of aging in human muscle reveals a common
  aging signature}.
\bjournal{PLoS Genetics}
\bvolume{2}
\bpages{1058--1069}.
\bptok{imsref}%
\end{barticle}
\endbibitem

\bibitem[\protect\citeauthoryear{Zahn et~al.}{2007}]{agemap}
\begin{barticle}[author]
\bauthor{\bsnm{Zahn},~\bfnm{J.~M.}\binits{J.~M.}},
  \bauthor{\bsnm{Poosala},~\bfnm{S.}\binits{S.}},
  \bauthor{\bsnm{Owen},~\bfnm{A.~B.}\binits{A.~B.}},
  \bauthor{\bsnm{Ingram},~\bfnm{D.~K.}\binits{D.~K.}},
  \bauthor{\bsnm{Lustig},~\bfnm{A.}\binits{A.}},
  \bauthor{\bsnm{Carter},~\bfnm{A.}\binits{A.}},
  \bauthor{\bsnm{Weeratna},~\bfnm{A.~T.}\binits{A.~T.}},
  \bauthor{\bsnm{Taub},~\bfnm{D.~D.}\binits{D.~D.}},
  \bauthor{\bsnm{Gorospe},~\bfnm{M.}\binits{M.}},
  \bauthor{\bsnm{Mazan-Mamczarz},~\bfnm{K.}\binits{K.}},
  \bauthor{\bsnm{Lakatta},~\bfnm{E.~G.}\binits{E.~G.}},
  \bauthor{\bsnm{Boheler},~\bfnm{K.~R.}\binits{K.~R.}},
  \bauthor{\bsnm{Xu},~\bfnm{X.}\binits{X.}},
  \bauthor{\bsnm{Mattson},~\bfnm{M.~P.}\binits{M.~P.}},
  \bauthor{\bsnm{Falco},~\bfnm{G.}\binits{G.}},
  \bauthor{\bsnm{Ko},~\bfnm{Mi~S.~H.}\binits{M.~S.~H.}},
  \bauthor{\bsnm{Schlessinger},~\bfnm{D.}\binits{D.}},
  \bauthor{\bsnm{Firman},~\bfnm{J.}\binits{J.}},
  \bauthor{\bsnm{Kummerfeld},~\bfnm{S.~K.}\binits{S.~K.}},
  \bauthor{\bsnm{III},~\bfnm{W.~H.~Wood}\binits{W.~H.~W.}},
  \bauthor{\bsnm{Zonderman},~\bfnm{A.~B.}\binits{A.~B.}},
  \bauthor{\bsnm{Kim},~\bfnm{S.~K.}\binits{S.~K.}} \AND
  \bauthor{\bsnm{Becker},~\bfnm{K.~G.}\binits{K.~G.}}
(\byear{2007}).
\btitle{{AGEMAP}: A gene expression database for aging in mice}.
\bjournal{PLoS Genetics}
\bvolume{3}
\bpages{2326--2337}.
\bptok{imsref}%
\end{barticle}
\endbibitem

\end{thebibliography}
\end{document}